\documentclass[useAMS,usenatbib]{mn2e}

\usepackage{amsmath,amssymb,bm,graphicx,epsfig,times}

\usepackage[normalem]{ulem} 
\usepackage{multirow}

\renewcommand{\vec}[1]{\bmath{#1}}
\newcommand{\corr}{{l}}
\newcommand{\cra}{_\mathrm{cr}} 
\newcommand{\dd}{\mathrm{d}} 
\newcommand\sfrac[2]{{\textstyle{\frac{#1}{#2}}}}
\newcommand{\bra}[1]{\left\langle #1\right\rangle}
\newcommand{\aver}[1]{\overline{#1}}

\newcommand{\RM}{\mathrm{RM}}

\newcommand{\cm}{\,{\rm cm}}    
\newcommand{\km}{\,{\rm km}}    
\newcommand{\m}{\,{\rm m}}      
\newcommand{\p}{\,{\rm pc}}     
\newcommand{\kpc}{\,{\rm kpc}}  
\newcommand{\GHz}{\,{\rm GHz}}  
\newcommand{\Hz}{\,{\rm Hz}}    
\newcommand{\MHz}{\, {\rm MHz}} 
\newcommand{\s}{\,{\rm s}}      
\newcommand{\yr}{\,{\rm yr}}    

\newcommand{\kms}{\km\s^{-1}}    

\newcommand{\mkG}{\,\mu{\rm G}} 

\newcommand{\GeV}{\,{\rm GeV}}  
\newcommand{\K}{\,{\rm K}}      

\newcommand{\fluc}{\,\sigma_{I}/I_0} 

\title{An observational test for correlations between cosmic rays and magnetic
fields}

\author[Stepanov et al.]{Rodion~Stepanov$^{1,2,3}$\thanks{E-mail:
rodion@icmm.ru (RS); anvar.shukurov@ncl.ac.uk (AS);
andrew.fletcher@ncl.ac.uk (AF); rbeck@mpifr-bonn.mpg.de (RB);
laporta@mpifr-bonn.mpg.de (LLP), taba@mpia.de (FT)},
Anvar~Shukurov$^1$, Andrew~Fletcher$^1$, Rainer~Beck$^{4}$,
\and
Laura~La~Porta$^{4,6}$, Fatemeh~Tabatabaei$^5$\\
$^1$School of Mathematics and Statistics, Newcastle University, Newcastle
upon Tyne, NE1 7RU, UK\\
$^2$Institute
of Continuous Media Mechanics, Academy of Sciences, Korolyov str. 1, Perm
614013, Russia\\
$^3$Department of Applied Mathematics and Mechanics, National Research Polytechnic University of Perm, Komsomolskii
Av. 29, 614990, Perm, Russia\\
$^4$Max-Planck Institut f\"ur Radioastronomie, Auf dem H\"ugel 69, Bonn
D-53121, Germany\\
$^5$Max Planck Institut f\"ur Astronomie, K\"onigstuhl 17,  Heidelberg D-69117, Germany\\
$^6$Institute of Geodesy and Geoinformation, University of Bonn, Nussallee 17, 53115 Bonn, Germany
}

\begin{document}

\date{Accepted  .... Received  ....; in original form ...}

\pagerange{\pageref{firstpage}--\pageref{lastpage}}
\pubyear{2009}

\maketitle

\label{firstpage}

\begin{abstract}
We derive the magnitude of fluctuations in total synchrotron intensity in the
Milky Way  and M33, from both observations and theory under various assumption
about the relation between cosmic rays and interstellar magnetic fields. Given
the relative magnitude of the fluctuations in the Galactic magnetic field
(the ratio of the rms fluctuations to the mean magnetic field strength)
suggested by Faraday rotation and
synchrotron
polarization, the observations are
inconsistent with local energy equipartition between cosmic rays and
magnetic fields. Our analysis of
relative
synchrotron intensity fluctuations
indicates
that the distribution of cosmic rays is nearly uniform at the scales of
the order of and exceeding $100\p$, in
contrast to
strong fluctuations in the interstellar magnetic field at those scales.
A conservative upper limit on the ratio of the
the fluctuation magnitude in the cosmic ray number density to its mean value is 0.2--0.4
at scales of order 100\,pc. Our results are consistent with a mild
anticorrelation between cosmic-ray and magnetic energy densities at these
scales,
in both the Milky Way and M33. Energy equipartition between cosmic rays and magnetic
fields may still hold, but at scales exceeding 1\,kpc.
Therefore, we suggest that equipartition estimates be applied to the
observed synchrotron intensity smoothed to a linear scale of kiloparsec order
(in spiral galaxies) to obtain
the cosmic ray distribution and a large-scale magnetic field. Then the resulting cosmic ray
distribution can be used to derive the fluctuating magnetic field strength from the data at the original
resolution. The resulting random magnetic field is likely to be significantly stronger than
existing estimates.
\end{abstract}


\section{Motivation and background}
The concept of energy equipartition between cosmic rays and magnetic fields
and similar assumptions such as pressure equality \citep{Longair:1994,BK05,AUAPV11}
are often used in the analysis and interpretation of radio astronomical
observations. This idea was originally suggested in order to estimate the magnetic field and cosmic ray
energies of the source \textit{as a
whole\/} \citep{Burbidge:1956a, Burbidge:1956b}, from a
measurement of the synchrotron brightness of a radio source. A physically attractive
feature of the equipartition state is that it approximately minimizes the
total energy of the radio source.

The energy density of cosmic rays is mainly
determined by their proton component, whereas the synchrotron intensity
depends on the number density of relativistic electrons. Therefore, in order
to estimate the magnetic field energy, an assumption
needs to be made about the ratio of the energy densities of the relativistic
protons and electrons; the often adopted value for this ratio is 100, as
suggested by Milky Way data \citep{BK05}. This ratio is adopted to be
unity in applications to galaxy clusters, radio galaxies and active objects
\citep{2002ARA&A..40..319C}.

However, more recently this concept has been extended to large-scale trends
in synchrotron intensity and to local
energy densities at sub-kiloparsec scales in well-resolved radio
sources, such as spiral galaxies
\citep[e.g.,][]{Beck:2005,B07,Chyzy:2008,2008A&A...490.1005T,FBSBH11}.
Another important application of the equipartition hypothesis, first suggested
by \citet{P66,P69,P79}, is to the hydrostatic equilibrium of the interstellar
gas. Here magnetic and cosmic ray pressures are assumed to be in a constant
ratio, in practice taken to be unity. This application appeals to
equipartition (or, more precisely, pressure equality) at larger scales of the
order of kiloparsec. The spatial relation between fluctuations in magnetic field and
cosmic rays is crucial for a proposed method to measure magnetic helicity in the ISM
\citep{Oppermann11,2010JETPL..90..637V}.

The physical basis of the equipartition assumption remains elusive. Since cosmic rays are
confined within a radio source by magnetic fields, it seems natural to expect
that the two energy densities are somehow related: if the magnetic field
energy density $\epsilon_B$ is smaller than that of the cosmic rays,
$\epsilon\cra$, the cosmic rays would be able to `break through' the magnetic
field and escape; whereas a larger magnetic energy density would result in the
accumulation of cosmic rays. Thus, the system is likely to be
self-regulated to energy equipartition, $\epsilon_B\approx\epsilon\cra$. A
slightly different version of these arguments refers to the equality of the two
pressures,\footnote{It is useful to carefully distinguish between what can be
called `pressure equality' and `pressure equilibrium': the former refers to
the case where magnetic fields and cosmic rays have equal pressures locally,
whereas the latter describes the situation where the sum of the two (or more)
pressure contributions does not vary in space.} giving
$\epsilon_B\approx\sfrac13\epsilon\cra$.

However plausible one finds these arguments, it is difficult to substantiate them.
In particular, models of cosmic ray confinement suggest that
the cosmic ray diffusion tensor depends on the \textit{ratio\/} $(\delta
B/B_0)^2$, where $\delta B$ is the magnitude of magnetic field fluctuations at
a scale equal to the proton gyroradius and $B_0$ is the mean magnetic field
\citep[e.g.,][]{BBDGP90}. The magnetic field strength can determine the streaming
velocity of cosmic rays via the Alfv\'en speed, but the theory of cosmic ray
propagation and confinement relates $\epsilon\cra$ to the intensity of cosmic
ray sources rather than to the local magnetic field strength. Despite their
uncertain basis, equipartition arguments remain popular as they provide
`reasonable' estimates of magnetic fields in radio sources, and also because
they often offer the only practical way to obtain such estimates.

Equipartition between cosmic rays and magnetic fields can rarely be tested
observationally. \citet{CW93} used $\gamma$-ray observations of the Magellanic
clouds to calculate the energy density of cosmic rays independently of the
equipartition assumption. They further calculated magnetic energy density from
radio continuum data at a wavelength of about $\lambda12\cm$. The resulting
magnetic energy density is two orders of magnitude larger than that of cosmic
rays, and \citet{CW93} argue that the discrepancy cannot be removed by
assuming a proton-to-electron ratio for cosmic rays different from the
standard value of 100 \citep[see, however,][]{P93}.
{ More recently, however, \citet{2012ApJ...759...25M}  analysed {\it Fermi} Large Area Telescope observations of the LMC \citep{2010A&A...512A...7A} and concluded that the equipartition assumption does not appear to be violated.}

An independent estimate of magnetic field strength can be obtained for
synchrotron sources of high surface brightness (e.g., active galactic nuclei)
where the relativistic plasma absorbs an observable lower-frequency part of
the radio emission (synchrotron self-absorption). Then the magnetic field strength
can be estimated from the frequency, the flux density and the angular size of
the synchrotron source at the turnover frequency \citep{S63,W63,SW68}.
\citet{SR77} and \citet{R94} concluded, from low-frequency observations of
compact radio sources whose angular size can be determined from interplanetary
scintillations, that there is no significant evidence of strong departures
from equipartition. In the sources with strong synchrotron self-absorption in
their sample, the total energy is within a factor of 10 above the minimum
energy.
\citet{OD08} observed, using VLBI, five young, very compact
radio sources to suggest
that magnetic fields in them are quite close to the equipartition value.
Physical conditions in spiral galaxies are quite different from those in
compact, active radio sources, and departures from equipartition by a factor
of several in terms of magnetic field strength would be quite significant in
the context of spiral galaxies.

Here we test the equipartition hypothesis using another approach
\citep[see also][]{2009IAUS..259...93S}.
We calculate the relative magnitude of fluctuations in synchrotron intensity using model
random magnetic field and cosmic ray distributions with a prescribed degree of
cross-correlations. When the results
are compared with observations, it becomes clear that local energy
equipartition is implausible as it would produce stronger fluctuations of the
synchrotron emissivity than are observed. Instead, the observed synchrotron intensity
fluctuations suggest weak variations in the cosmic ray number
density or an anticorrelation between cosmic rays and magnetic fields,
perhaps indicative of pressure equilibrium. We conclude that local energy
equipartition is unlikely in spiral galaxies at the integral scale of the
fluctuations, of order 100\p. We discuss the dynamics of cosmic rays to argue
in favour of equipartition at larger scales of order 1\,kpc, comparable to the
scale of the mean magnetic field and to the cosmic-ray diffusion scale.

{ The paper is organized as follows. In Section \ref{MFIMF} we
discuss the relative strengths of the mean and fluctuating magnetic field in the Milky
Way and M33. In Section~\ref{data} we
use observational data to estimate
the magnitude of synchrotron intensity variations at high galactic latitudes
in the Milky Way and in the outer parts of M33.
Theoretical models for synchrotron intensity fluctuations, allowing for controlled levels of cross-correlation between the magnetic field and cosmic ray distributions, are developed analytically in Section \ref{SIF} and numerically in Section \ref{MFCRM}: readers who are interested only in our results may wish to skip these rather mathematical sections.
Section \ref{SR} presents an interpretation of the observational data
in terms of the theoretical models; here we estimate the cross-correlation coefficient
between magnetic and cosmic-ray fluctuations. In Section~\ref{SDCR} we briefly
discuss cosmic ray propagation models from the viewpoint of relation between
the cosmic ray and magnetic field distributions. Our results are discussed in
Section~\ref{Disc} and Appendix~\ref{app} contains the
details of some of our calculations.}

\section{The magnitude of fluctuations in interstellar magnetic fields}\label{MFIMF}
The ratio of the fluctuating-to-mean synchrotron intensity in the interstellar
medium (ISM) is sensitive to the relative distributions of cosmic ray
electrons and magnetic fields and hence to the extent that energy
equipartition may hold locally: the synchrotron emission will fluctuate
strongly if equipartition holds pointwise, i.e., if the number density of
cosmic ray electrons is increased where the local magnetic field is stronger.
(We assume that cosmic ray electrons and heavier relativistic particles are similarly
distributed -- see Section~\ref{Disc} for the justification.)

Interstellar magnetic fields are turbulent, with the ratio of the
random magnetic field to its mean component known from observations of Faraday
rotation, independently of the equipartition assumption. Denoting the standard
deviation of the turbulent magnetic field by $\sigma_b^2=\aver{B^2}-B_0^2$
and the mean field strength as $B_0=|\aver{\vec{B}}|$, where bar denotes
appropriate averaging (usually volume or
line-of-sight averaging), the relative fluctuations in magnetic field
strength in the Solar vicinity of the Milky Way is estimated as
\citep{RSS88,OS93,BBMSS96}
\begin{equation}\label{ratioB}
\delta_b^2=\left(\frac{\sigma_b}{B_0}\right)^2\simeq3\mbox{--}10.
\end{equation}

Similar estimates result from radio observations of nearby spiral galaxies
where the degree of polarization of the integrated emission at 4.8\,GHz is a
few per cent, with a range $p\simeq0.01$--$0.18$ \citep{2009ApJ...693.1392S}.
These data are affected by beam depolarization, so they only give
upper limits for $\delta_b^2$. More typical values of the fractional polarisation in spiral galaxies
are $p=0.01$--0.05 within spiral arms and 0.1 on average. The degree of
polarization at short wavelengths, where Faraday rotation is negligible, can be
estimated as \citep{1966MNRAS.133...67B,1998MNRAS.299..189S}
\begin{equation}\label{burn1}
p=p_0 \frac{B_{0\perp}^2}{B_{0\perp}^2+\tfrac{2}{3}\sigma_b^2}
\end{equation}
where $B_{0\perp}$
is the strength of the large-scale magnetic field in the sky plane, {the intrinsic degree of polarisation}
$p_0\approx0.75$, and the random magnetic field is assumed to be isotropic,
$\aver{b_\perp^2}=\tfrac23\sigma_b^2$. This yields
\begin{equation}\label{burn}
\delta_b^2\simeq\tfrac32\left(\frac{p_0}{p}-1\right)\ga4
\quad\mbox{for}\quad p<0.2\,
\end{equation}
in a good agreement with {the estimate \eqref{ratioB} for} the Milky Way data obtained from Faraday rotation
measures.
For $p=0.05$--0.1, we obtain $\delta_b^2\simeq10$--20.

It is important to note that Eq.~(\ref{burn}) has been obtained assuming
that the cosmic ray number density $n\cra$ is uniform, so that all the beam
depolarization is attributed solely
to the fluctuations in magnetic field. Under local equipartition,
$n\cra\propto B^2$, \citet[their Eq.~(28)]{1998MNRAS.299..189S} calculated the degree of
polarization at short wavelengths to be
\[
p = p_0\frac{1+\tfrac73\delta_b^2}{1+3\delta_b^2+\tfrac{10}{9}\delta_b^4}.
\]
As might be expected, this expression leads to a smaller $\delta_b$ for a given
$p/p_0$ than Eq.~(\ref{burn1}):
\begin{equation}\label{delta_b_eq}
\delta_b^2\approx \frac{2 p_0}{p}
\quad
\mbox{for}\quad \frac{p}{p_0}\ll1,
\end{equation}
so that
\[
\delta_b^2\simeq15\quad\mbox{for} \quad p=0.1.
\]
Since local equipartition between cosmic rays and magnetic fields maximizes
beam depolarization, this is clearly a lower estimate of $\delta_b$.

\subsection{{Anisotropic fluctuations}}\label{AF}
The above estimates apply to statistically isotropic random magnetic fields.
However, the random part of the interstellar magnetic field can be expected to
be anisotropic at scales of order 100\,pc, e.g., due to shearing by the
galactic differential rotation, streaming motions and large-scale compression.
Synchrotron emission arising in an
anisotropic random magnetic field is polarized \citep{L80,1998MNRAS.299..189S}
and the resulting net polarization, from the combined random and mean field,
can be either stronger or weaker than in the case of an isotropic random field
depending on the sense of anisotropy relative to the orientation of the mean
magnetic field. Note that the anisotropy of MHD turbulence resulting from the nature of the
spectral energy cascade \citep[][and references therein]{GS95,LGS07,GNNP00} is
important only at much smaller scales.

The case of M33 provides a suitable illustration of the refinements required
if the anisotropy of the random magnetic field is significant.
\citet[][their Table~1]{2008A&A...490.1005T} obtained
{integrated}
fractional polarization of about 0.1
at $\lambda3.6\cm$. Using Eq.~(\ref{burn}), this yields
$\delta_b^2\simeq10$, whereas Eq.~(\ref{delta_b_eq}) leads to
$\delta_b^2\simeq4$, consistent with their equipartition estimates
$\sigma_b\simeq6\mkG$ and $B_0\simeq2.5\mkG$. However, their analysis of
Faraday rotation between $\lambda3.6,\ 6.2$ and $20\cm$ suggests a weaker
regular magnetic field, $B_0\simeq1\mkG$, leading to $\delta_b^2\simeq40$ if
$\sigma_b\simeq6\mkG$.

{The latter estimate for $B_0$ is more reliable since the degree of
polarization leads to an underestimated $\delta_b$ if magnetic field
is anisotropic.}
\citet[][their Eq.~(19)]{1998MNRAS.299..189S}
have shown that the degree of polarization at short wavelengths in a partially
ordered, anisotropic magnetic field  is given by
\[
p=p_0\frac{1+\delta_{by}^2(1-\alpha_b^2)}{1+\delta_{by}^2(1+\alpha_b^2)},
\]
where
{the $(x,y)$-plane is the plane of the sky with the $y$-axis aligned
with the large-scale magnetic field, i.e., $B_y=B_0$ and $B_x=0$;
we further defined}
$\delta_{by}^2=\sigma_{by}^2/B_0^2$ and likewise for $\delta_{bx}$, and
introduced $\alpha_b^2=\sigma_{bx}^2/\sigma_{by}^2$ ($<1$) as a measure of the
anisotropy of $\vec{b}_\perp$. This approximation is relevant to spiral
galaxies where the mean magnetic field is predominantly azimuthal (nearly
aligned with the $y$-axis of the local reference frame used here) and the
anisotropy in the random magnetic field is produced by the rotational shear,
$\sigma_{by}>\sigma_{bx}$. For $\delta_{by}^2\gg1$, this yields, for $p=0.1$,
\begin{equation}\label{alphab}
\alpha_b^2\approx\frac{p_0-p}{p_0+p}\approx0.8.
\end{equation}
Thus, a rather weak anisotropy of the random magnetic field can produce
$p\simeq0.1$ and this allows us to
reconcile the different estimates of $\delta_b$ obtained from the degree of polarization
and Faraday rotation in M33.

The required anisotropy can readily be produced by the galactic differential
rotation. Shearing of an initially isotropic random magnetic field
{by rotation (directed along the $y$-axis)}
leads, within one eddy turnover time, to an increase of its azimuthal component to
\[
\sigma_{by}\simeq\sigma_{bx}\left(1-r\frac{\dd\Omega}{\dd r}\,
        \frac{\corr}{v}\right),
\]
where $\Omega(r)$ is the angular velocity of the galactic rotation (with the
rotational velocity along the local $y$-direction and $r$ the galactocentric
radius) and $\corr$ and $v$ are the correlation length\footnote{The correlation
length $l$ is also known as the integral scale and is defined as the integral of
the variance-normalized autocorrelation function of a random variable. The
typical linear size, or diameter, of a turbulent cell is $2l$.} and r.m.s.\ speed of the
interstellar turbulence, so that $\corr/v$ is the lifetime of a turbulent
eddy. This leads to
\[
\alpha_b\simeq
        \left(1-r\frac{\dd\Omega}{\dd r}\,\frac{\corr}{v}\right)^{-1}
\simeq 1-\frac{\corr V_0}{R_0 v},
\]
where the last equality is based on the estimate $r\dd\Omega/\dd r\simeq
-V_0/R_0$, with $V_0=107\kms$ and $R_0=8\kpc$ being the parameters of Brandt's
approximation to the rotation curve of M33 \citep{RWL76}. With
$\corr=0.1\kpc$ and $v=10\kms$
\citep[values typical of spiral galaxies -- e.g., Sect.~VI.3 in][]{RSS88},
we obtain $\alpha_b^2\simeq0.8$ , in
perfect agreement with the degree of anisotropy
required by {Eq.~\eqref{alphab}} to explain the observations of \citet{2008A&A...490.1005T}.

\subsection{{Summary}}
To conclude, a typical value of the relative strength of the random magnetic
field in spiral galaxies is, at least,
\begin{equation}\label{delta_b_st}
\delta_b^2\simeq10.
\end{equation}
This estimate refers to the correlation scale of interstellar turbulence, $\corr\simeq50$--$100\p$.
{The correlation scale will be introduced in Section~\ref{SIF}, but here we stress
that this estimate refers to the larger scales in the turbulent spectrum}.
Higher values, $\delta_b^2\simeq40$ are perhaps more plausible,
{especially within spiral arms,} but our results are not very sensitive
to this difference (see Fig.~\ref{I1b} and Section~\ref{SIF}).
{The estimate of $\delta_b$ in the Milky Way refers to the
solar vicinity, i.e., to a region  between major spiral arms where the degree of polarization
is higher than within the arms and, correspondingly, $\delta_b$ is larger.
Consistent with this, our analysis of the observed synchrotron fluctuations in
Section~\ref{data} is for high Galactic latitudes and the
outer parts of M33 where the influence of the spiral arms is not strong. Overall,
$3\la\delta_b^2\la40$ appears to be a representative range for spiral galaxies,
excluding their central parts.}

\section{Synchrotron intensity fluctuations derived from observations of the Milky Way and M33}\label{data}
In this section we estimate the relative level of synchrotron intensity fluctuations
from observations of the Milky Way and the spiral galaxy M33. An ideal data
set for this analysis should: (i)~resolve the fluctuations at their largest
scale, (ii)~only include emission from the ISM and not from discrete sources
such as AGN and stars, (iii)~not be dominated by structures that are large and
bright due to their proximity, such as the North Polar Spur, (iv)~be free of
systematic trends such as arm-interarm variations or vertical
stratification. The data should allow the ratio
\begin{equation}
\delta_I=\frac{\sigma_I}{I_0} \label{deltaI},
\end{equation}
{where $\sigma_I$ and $I_0$ are the standard deviation
and the mean value of synchrotron intensity in a given region,} to be calculated separately in arm and inter-arm regions or at low and high
latitudes as $I_0$ differs between these regions.
Regarding item (i) above, we
note that a turbulent cell of $100\p$ in size subtends the angle of about
$6\degr$ at a distance of $1\kpc$. Furthermore,  most useful for our purposes
are long wavelengths where the contribution of thermal radio emission is
minimal. Unfortunately, ideal data satisfying all these criteria do not exist;
we therefore use several radio maps, where each map possesses a few of the
desirable properties listed above and collectively they have them all.

{The Milky Way maps that we use contain isotropic emission from faint, unresolved extra-galactic sources and the cosmic microwave background. The contribution of the extragalactic sources to the brightness temperature of the
total radio emission of the Milky Way is estimated by}
\citet{LMOP87}
{as
$T_\mathrm{e}\simeq50\K(\nu/150\MHz)^{-2.75}$, which amounts to
about $10^4\K$ and $3\K$ at the frequencies $\nu=22\MHz$ and $408\MHz$,
respectively. The $3\K$ temperature of the cosmic microwave background should also be taken into account at $408\MHz$. For comparison, the respective total values of the radio brightness temperature
near the north Galactic pole are about $3\times10^4\K$ and  $20\K$ at $22\MHz$ and $408\MHz$, respectively.
In our estimates of $\delta_I$ obtained below,
we have not subtracted this contribution from $I_0$. Thus, our estimates of $\delta_I$ are
conservative, and more realistic values might be about 40\% larger at both $22\MHz$ and $408\MHz$.
The observations of M33 use a zero level that is set at the edges of the observed area of the sky; since this zero level includes the CMB and unresolved extra-galactic sources $\delta_I$ is unaffected by these components.}

\subsection{The data}
\subsubsection{The 408\,MHz all-sky survey} \label{408MHz}
The survey of \citet{Haslam:1982} covers the entire sky at a resolution of
$51\arcmin$ {(about $50\p$ for a distance of $1\kpc$)} and with an estimated noise level of about $0.67$~K. Synchrotron
radiation is the dominant contribution to emission at
the survey's wavelength of $\lambda$74\,cm. The brightest structures in the
map shown in Fig.~\ref{sky408all}a are the Galactic plane and several arcs due
to nearby objects, especially the North Polar Spur.

We expect that results useful for our purposes arise at the angular scale of
about $6\degr$ in all three Milky Way maps (i.e., the angular size of a
turbulent cell at a 1\,kpc distance), whereas larger scales  { reflect} regular
spatial variations of the radio intensity.

\subsubsection{The 408\,MHz all-sky survey, without discrete sources}
\citet{LaPorta:2008} removed the strongest discrete sources from the data of
\citet{Haslam:1982} using a two-dimensional Gaussian filter. We compared the
results obtained from the original $408\MHz$ survey with those from this map
to show that the effect of point sources on our results is negligible.

\subsubsection{The 22\,MHz part-sky survey} \label{22MHz}
\citet{Roger:1999} produced a map, shown in Fig.~\ref{sky22}a, of about $73\%$
of the sky at $\lambda13.6$\,m, between declinations $-28\degr$ and $+80\degr$
at a resolution of approximately $1\degr\times 2\degr$ and an estimated noise
level of $5$~kK. The emission is all synchrotron radiation, but  H\,{\sc ii}
regions in the Galactic plane absorb some background emission at this low
frequency. However, we are most interested in regions away from the Galactic
plane, so our conclusions are not affected by the absorption in the H\,{\sc
ii} regions. The brightest point sources were removed by \citet{Roger:1999} as
they produced strong sidelobe contamination in the maps: this accounts for the
four empty rectangles in Fig.~\ref{sky22}a.

\subsubsection{The 1.4\,GHz map of M33}
The nearby, moderately inclined, spiral galaxy M33 was observed at
$\lambda$21\,cm by \citet{TKB07}, using the VLA and Effelsberg
 telescopes, at a resolution of $51\arcsec$, or about $200\p$ at the
distance to M33 of $840\kpc$. The noise level is estimated to be $0.07$~mJy/beam.
The resolution is sufficient to resolve arm and inter-arm regions, but is at
the top end of the expected scale of random fluctuations due to turbulence.
The beam area includes a few (nominally, four) correlation cells of the
synchrotron intensity fluctuations. The emission is a mixture of thermal and synchrotron
radiation. The overall thermal fraction is estimated to be $18\%$ but it is
strongly enhanced in large H\,{\sc ii} regions and spiral arms
\citep{TBKKBGM07} whereas the synchrotron emission comes from the whole disc.
The radio map used here is shown in Fig.~\ref{M33}a. The spiral pattern in
notably weak in total synchrotron intensity, so the map appears almost
featureless. This makes this galaxy especially well suited for our analysis
since we are interested in quasi-homogeneous random fluctuations of the radio
intensity. Nevertheless, systematic trends are noticeable in this map and we
discuss their removal in Section~\ref{ssM33}.

\subsection{Statistical parameters of the synchrotron intensity fluctuations}
For the three Milky Way data sets of Sections~\ref{408MHz}--\ref{22MHz}, we
calculated the mean $I_0$ and standard
deviation $\sigma_I$ of the synchrotron intensity $I$ at each point in the map.
In each case, the data were smoothed with a Gaussian of an angular width $a$,
resulting in the local mean intensity at the scale $a$, which we denote
$I_{0a}$:
\begin{equation}\label{aver}
I_{0a} =S_a^{-1}\int\!\!\!\int I(l',b')
   \exp\left(-\theta^2/a^2\right) \cos{b}\,\dd l'\,\dd b',
\end{equation}
where integration extends over the data area,
$\vec{r}=(1,l,b)$ is the position vector on the unit sky sphere, with $l$ and
$b$ the Galactic longitude and latitude (confusion with the small-scale
magnetic field, denoted here $\vec{b}$, should be avoided),
$\theta=\arccos(\vec{r}\cdot\vec{r}')$
is the angular separation between $\vec{r}$ and $\vec{r}'$,
\[
S_a(l,b) =\int\!\!\!\int
   \exp\left(-\theta^2/a^2\right) \cos{b}\,\dd l'\,\dd b',
\]
is the averaging area,
and the integration extends over the whole area of the sky available in
a given survey. The standard deviation $\sigma_{Ia}$  of radio intensity at a
given position $(l,b)$ at a given scale $a$ is calculated as
\[
\sigma^2_{Ia}(l,b)=\bra{I^2}_a-\bra{I}_a^2,
\]
where angular brackets denote
{spatial}
averaging as defined  in Eq.~(\ref{aver}).

In the case of M33, we selected nine areas which avoid the brightest
H\,{\sc ii} regions and whose radio continuum emission is thus likely to be dominated by synchrotron radiation. Each area encompasses several beams and $\delta_I$ was
calculated for each area, using the mean value and the standard deviation of
$I$ among all the pixels in the field obtained after removing regular trends (see Section~\ref{ssM33}).

\begin{figure}
\begin{center}
 \includegraphics[width=0.46\textwidth]{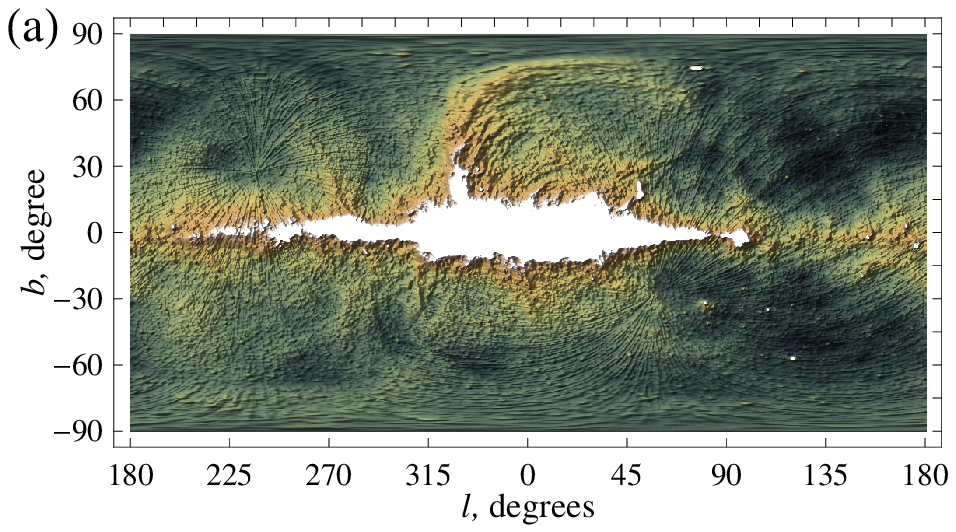}\\
 \includegraphics[width=0.35\textwidth]{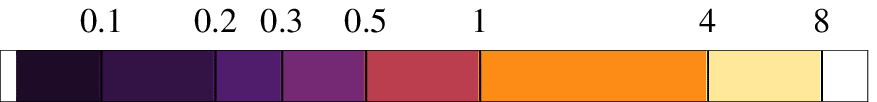}\\
 \includegraphics[width=0.46\textwidth]{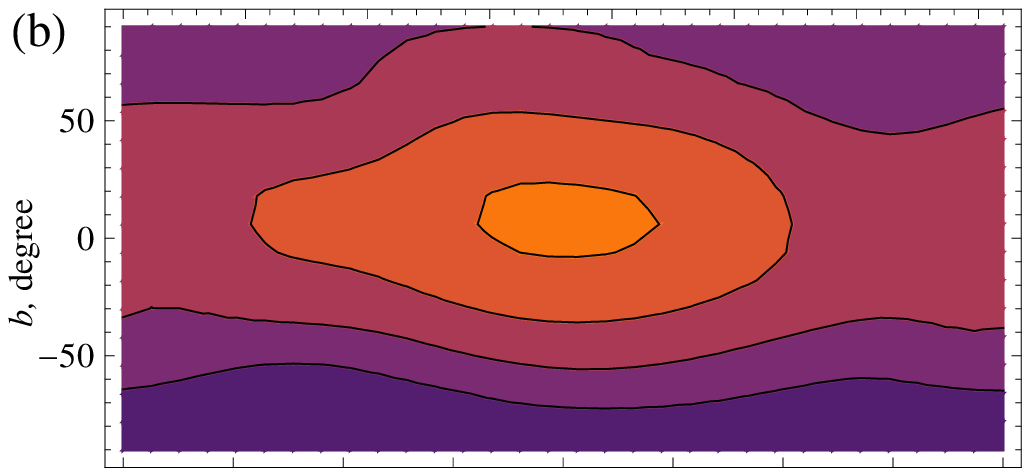}\\
 \includegraphics[width=0.46\textwidth]{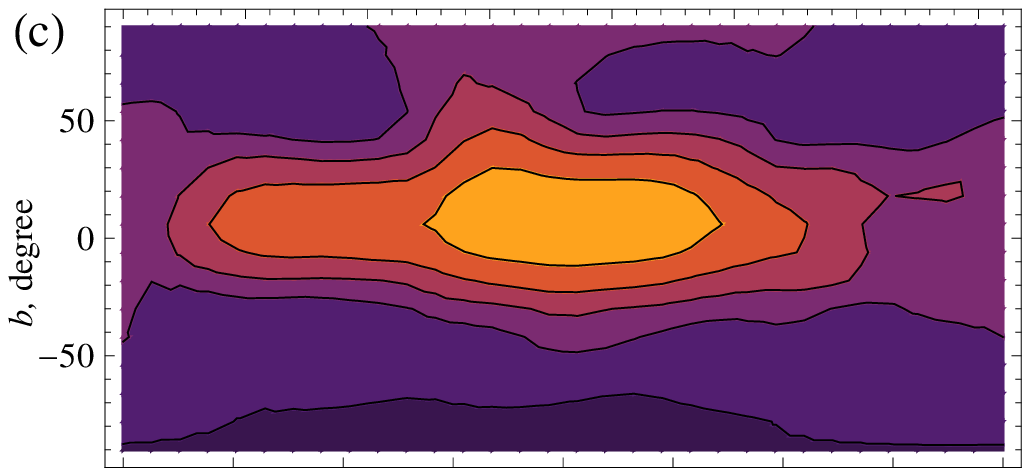}\\
 \includegraphics[width=0.46\textwidth]{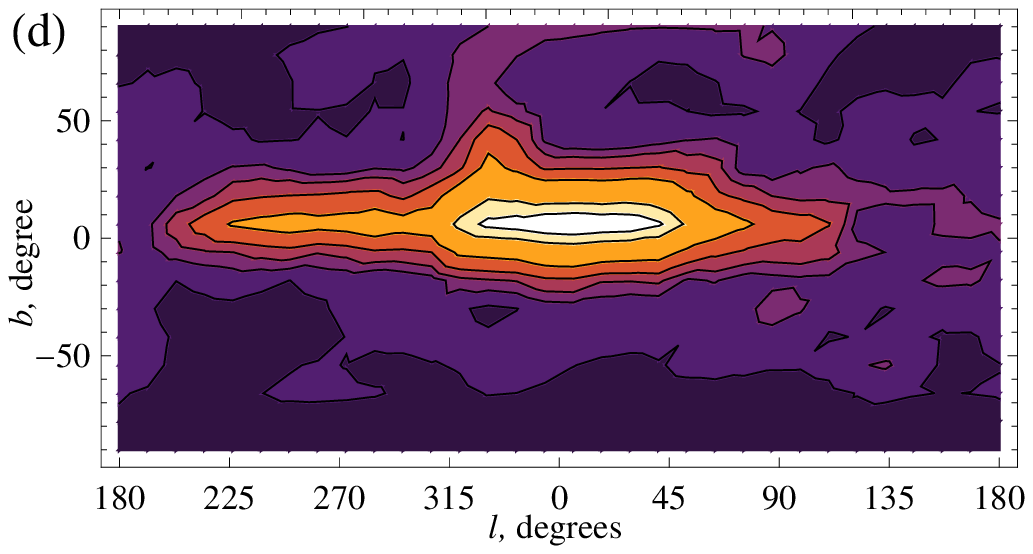}
\end{center}
 \caption{\label{sky408all} { (a):} The $408\MHz$ all-sky map of the total
synchrotron intensity \citep{Haslam:1982}, with the Galactic disc area
($I>52\,$K) blanked out. The lower panels show the magnitude of the relative
fluctuations of the synchrotron intensity, $\delta_I=\fluc$, at various
scales,
{with the colour bar shown between Panels (a) and (b)}:
{ (b)} $a=30\degr$, { (c)}  $a=15\degr$ and { (d)}
$a=7\degr$. The latter scale is about the angular size of a turbulent
cell ($2\corr_\varepsilon=100\p$) seen at a distance $1\kpc$.}
\end{figure}

\subsubsection{The 408\,MHz survey} \label{subsubsec:408}
To reduce the influence of the Galactic disc, where the
structure in the radio maps is mainly due to systematic arm-interarm
variations and localized radio sources such as supernova remnants, the
original intensity data were truncated at 52\,K (1\% of the maximum and 167\%
of the r.m.s.\ value of $I$) {: this blanking only affects emission at Galactic latitudes $|b|<20\degr$}. The resulting sky distribution of the relative
radio intensity fluctuations $\sigma_{Ia}/I_{0a}$ is shown in
Fig.~\ref{sky408all} for a selection of averaging scales, $a=30\degr, 15\degr$
and $7\degr$ ($a$ corresponds to the radius rather than the diameter
of the region).

While panels (b) and (c) in Fig.~\ref{sky408all} reflect mainly systematic
trends in radio intensity, we expect that panel (d) is dominated by the
turbulent fluctuations. In particular, the $a=7\degr$ map shows a much weaker  variation with Galactic latitude than those at larger scales,
for $|b|\ga30\degr$. We note that the correlation scale
of the synchrotron intensity fluctuations obtained by \citet{DS87} is $8\degr$.

{At $|b|\gtrsim20\degr$, the maximum pathlength through the synchrotron layer is about $h_{\epsilon}/\sin{b}\simeq 6\kpc$, where the synchrotron scale height $h_{\epsilon}\simeq 1.8\kpc$ \citep{Beuermann:1985}. With an angular resolution of about $1\degr$ the linear beamwidth is about $100\p$ at most. Give that the correlation length is about $l_{\epsilon}\simeq 50\p$ (see Section \ref{subsec:corr}) the beam encompasses one synchrotron cell at most.}

Contours outside the Galactic disc in Fig.~\ref{sky408all}d, $|b|\ga30\degr$ give
\begin{equation}\label{fluct}
\delta_I=0.1\mbox{--}0.2.
\end{equation}

Results obtained from the $408\MHz$ map with point sources subtracted
differ insignificantly from those obtained using the original map.

\begin{figure}
\begin{center}
 \includegraphics[width=0.47\textwidth]{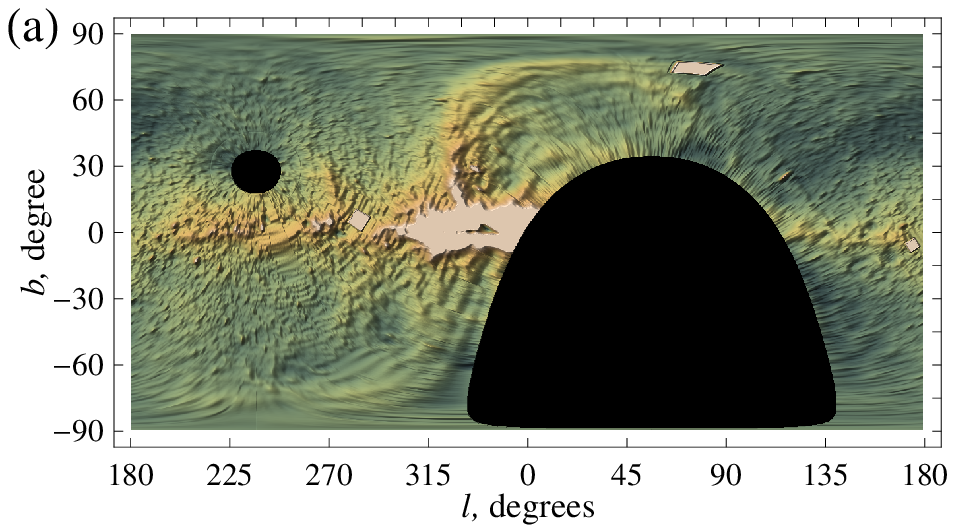}
 \includegraphics[width=0.35\textwidth]{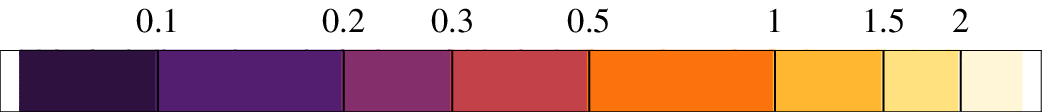}
 \includegraphics[width=0.47\textwidth]{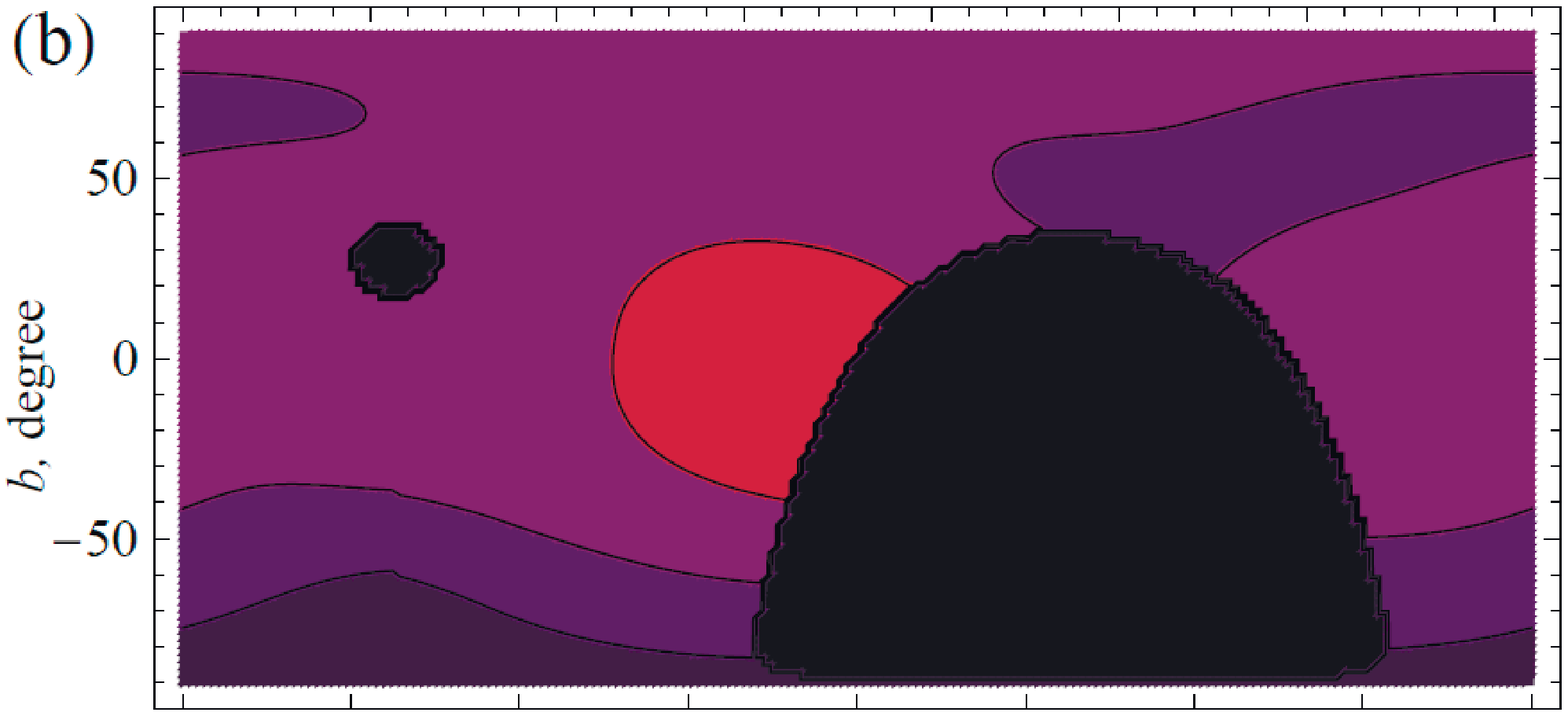}
 \includegraphics[width=0.47\textwidth]{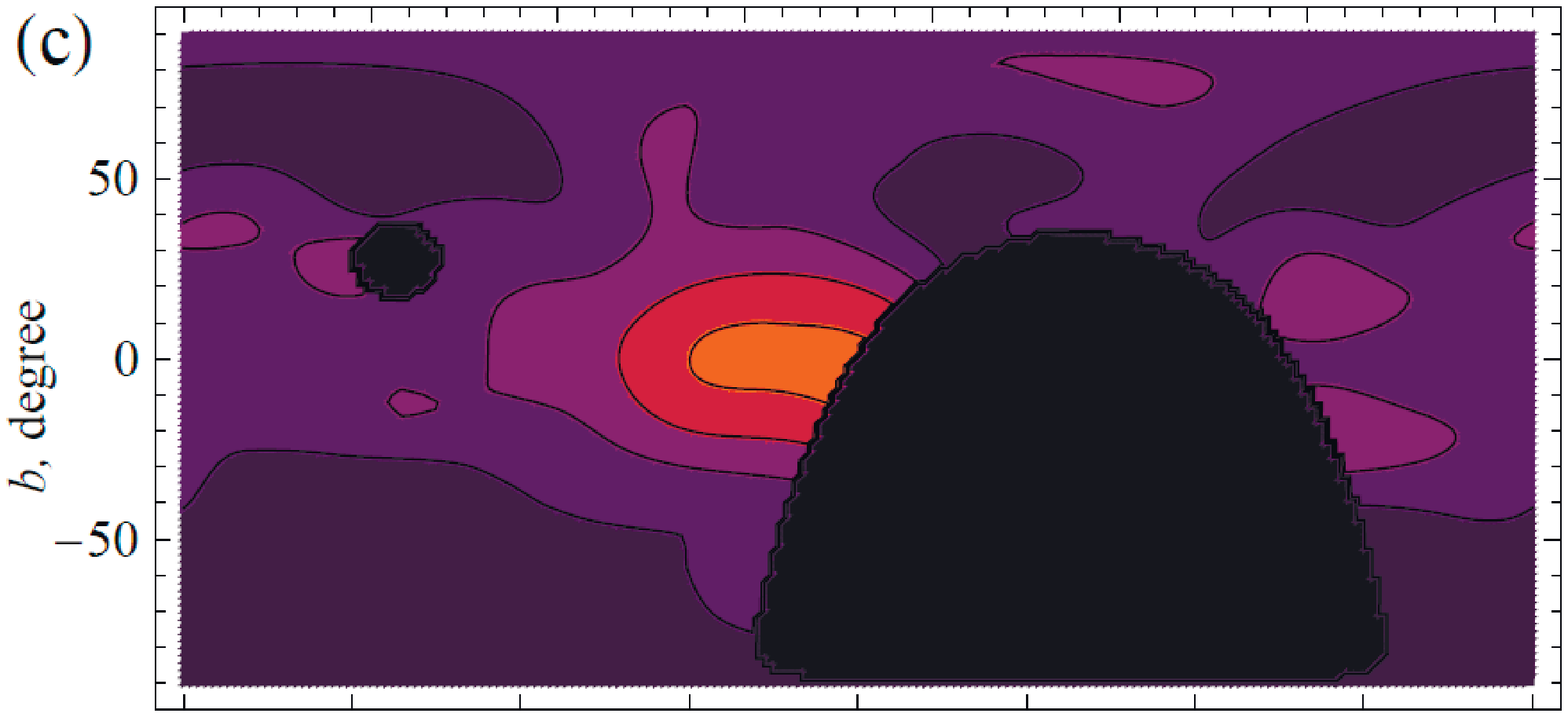}
 \includegraphics[width=0.47\textwidth]{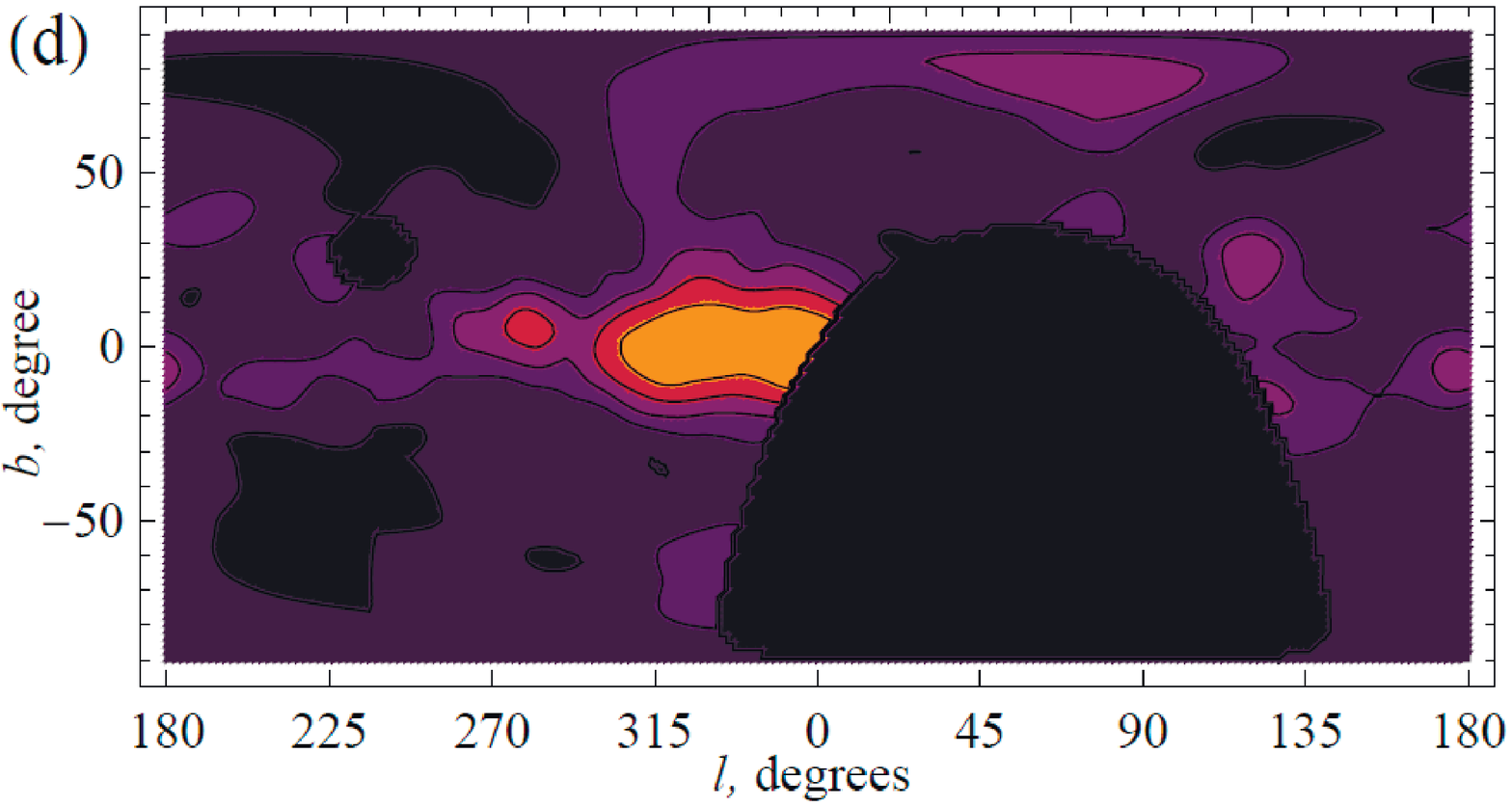}
\end{center}
\caption{ \label{sky22} As in Fig.~\ref{sky408all} but for the 22\,MHz survey
\citep{Roger:1999} with the Galactic disc area
($I>10^5\,$K) blanked out.}
\end{figure}

\subsubsection{The 22\,MHz partial sky survey}

Contours of the relative intensity fluctuations obtained from the $22\MHz$ map
are shown in Fig.~\ref{sky22}. As in Fig.~\ref{sky408all}, averaging over
scales $a=30\degr$ and $15\degr$ reveals the large-scale structure clearly
visible in the original data. However, results at $a=7\degr$ show much less of
such structure, and the statistically homogeneous part of the sky in this
panel includes the same contours of 0.1 and 0.2 as in Fig.~\ref{sky408all},
with the value of 0.3 confined to the bright ridges seen in Fig.~\ref{sky22}a.
Thus, the $22\MHz$ data are in a good agreement with the values for $\delta_I$
obtained from the 408 MHz data. This suggests a weak frequency dependence of the relative
synchrotron intensity fluctuations $\delta_I$.

\begin{figure}
\centering
 \includegraphics[width=0.44\textwidth]{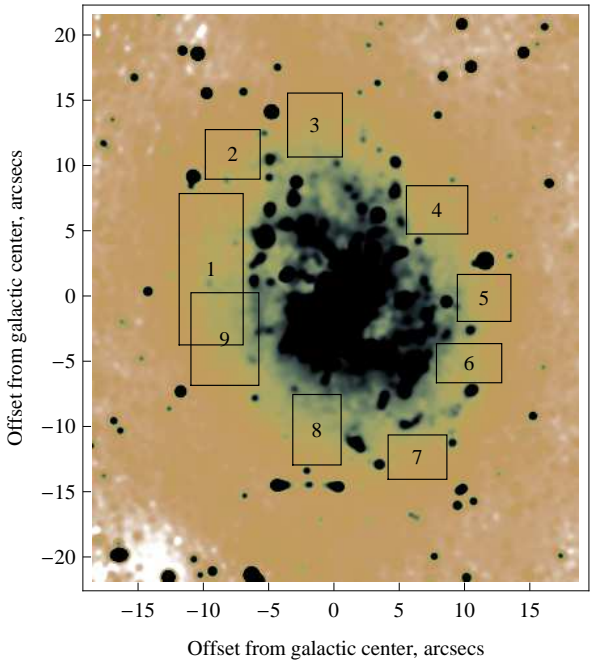}
 \caption{\label{M33} The $1.4\GHz$ radio map of M33 \citep{TKB07}, with the
rectangular fields used in our analysis shown. For orientation, we note that
the size of Field 3 is $7'\times 7'$ equivalent to $1.6\times 1.9\kpc^2$.}
\end{figure}

\begin{table}
\caption{\label{tab}Relative radio intensity fluctuations $\sigma_I/I_0$
in M33 with systematic trends of various orders subtracted. The first column
gives the field number as specified in Fig.~{\protect{\ref{M33}}}, and the
next three columns show the relative fluctuations of radio intensity, with
$\sigma^{(m)}_I$ the standard deviation of $I$ within a given field,
obtained with a trend of order $m$
  subtracted
{(the mean value of the trend vanishes across each field)}: $m=0$ corresponds to
the original data, $m=1$, to a linear trend, and $m=2$, to a quadratic trend
in the angular coordinates. The last column shows the mean value of the radio
intensity in each field.}
\begin{center}
\begin{tabular}{ccccc}
 \hline
Field No. & $\sigma^{(0)}_I/I_0$ & $\sigma^{(1)}_I/I_0$
                            & $\sigma^{(2)}_I/I_0$ & $I_0$ [$\mu$Jy/beam]\\
 \hline
 1 & \text{0.30 } & \text{0.18 } & \text{0.14 } & 677 \\
 2 & \text{0.34 } & \text{0.22 } & \text{0.19 } & 545 \\
 3 & \text{0.31 } & \text{0.15 } & \text{0.11 } & 609 \\
 4 & \text{0.28 } & \text{0.18 } & \text{0.17 } & 606 \\
 5 & \text{0.38 } & \text{0.17 } & \text{0.13 } & 672 \\
 6 & \text{0.35 } & \text{0.15 } & \text{0.12 } & 795 \\
 7 & \text{0.30 } & \text{0.16 } & \text{0.14 } & 565 \\
 8 & \text{0.32 } & \text{0.07 } & \text{0.07 } & 902 \\
 9 & \text{0.38 } & \text{0.14 } & \text{0.11 } & 810\\
\hline
Mean&\text{0.33 } & \text{0.16 } & \text{0.13 } & 687\\
  \hline
\end{tabular}
\end{center}
\end{table}

\subsubsection{M33}\label{ssM33}
Our analysis for the Milky Way has a potential difficulty that long lines of
sight might make it impossible to separate the contributions to $\sigma_I$
from small-scale (random) and large-scale (systematic) variations of the
synchrotron emissivity. Therefore, we consider also the nearby galaxy M33 seen
nearly face-on (inclination $56\degr$). To avoid excessive contribution from
large-scale variations due to the spiral pattern and the radial gradient in $I$, we
selected  areas free
of strong star formation in the outer regions of the galaxy
disc in Fig.~\ref{M33}. The areas of the rectangular fields
chosen range from $1.9\times 1.2\kpc^2$ to $1.9\times 4.7\kpc^2$.

The fields are big enough to make gradients in the mean quantities
significant; in particular, the non-thermal disc of M33 has a strong radial
gradient in radio intensity \citep{TBKKBGM07}, so we
subtract regular trends from the values
of $I$. We fitted first- or second-order polynomials to
$I(l,b)$ in $l$ and $b$ in each field and calculated $\delta_I$ after
subtracting the trends {with
{vanishing} mean value} from the original data. Results are shown in
Table~\ref{tab}, with $\sigma_I^{(m)}$ denoting the standard deviation of the
radio intensity obtained upon the subtraction of a polynomial of order $m$
in the angular coordinates.
The mean value of synchrotron intensity $I_0$ was calculated for each field.

We note that $\sigma_I^{(0)}$ (the standard deviation of $I$ in the original
data) is noticeably larger than $\sigma_I^{(1)}$ and $\sigma_I^{(2)}$. Thus,
the large-scale trends contribute significantly to the intensity variations.
On the other hand, $\sigma_I^{(1)}$ and $\sigma_I^{(2)}$ have rather similar
magnitudes of
{about}
0.15 (and $\sigma_I^{(1)}>\sigma_I^{(2)}$ as expected), so
that they can be adopted as an estimate of $\sigma_I$ corrected for the
large-scale trends. We use the value of $\sigma_I^{(2)}/I_0$ averaged over the
nine fields explored as the best estimate for $\delta_I$.

\begin{figure}
\bigskip
\includegraphics[width=0.4\textwidth]{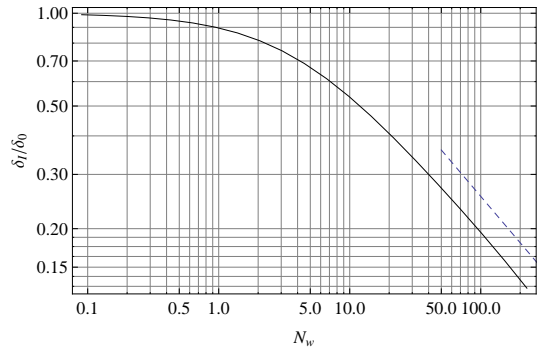}
\bigskip
\caption{\label{smallNW} The relative fluctuations in synchrotron intensity
$\delta_I$  as a function of $N_W$, the number of synchrotron correlation
cells across the beam area, with $\delta_I$ normalized by $\delta_0$, the
relative fluctuations obtained along a single line of sight, i.e., for $N_W\to0$.
The asymptotic dependence $\delta_I=2\delta_0 N_W^{-1/2}$ { (dashed line)} emerges only for
$N_W\ga20$--$30$. The calculation assumed a Gaussian beam, with the full width
at half-maximum (FWHM) taken for the beam diameter.}
\end{figure}

However, $W\approx100\p$ (half-width at half maximum of the Gaussian beam) in the
observations of M33 used here. {Assuming the synchrotron correlation length $l_{\epsilon}=50\p$} the beam area encompasses $N_W\approx4$ correlation cells.
{Therefore, to make the M33 results (especially $\delta_I$) comparable to those of the Milky Way. we have to reduce them to a common pathlength and number of synchrotron correlation cells within the beam. We recall that the beams at 408 MHz and 22 MHz only cover at most one synchrotron cell (see Section \ref{subsubsec:408}), so we have $N_W=1$ in the Milky Way.}
For the small value of $N_W$ in the high-resolution observations of M33,
the dependence of $\delta_I$ differs significantly from its
asymptotic form $\delta_I\propto N_W^{-1/2}$. Figure~\ref{smallNW} shows the dependence of $\delta_I$ on
$N_W$ obtained for a model synchrotron-emitting system described in detail in
Section~\ref{MFCRM}: $\delta_I$ only weakly depends on $N_W$ for small $N_W$.
Reduced to a single line of sight, the synchrotron intensity fluctuations in M33 then
correspond to $\delta_I=0.13/0.7=0.2$ if the synchrotron correlation length is
$\corr_\varepsilon=50\p$ (i.e., $N_W=4$). We also recall that the typical pathlength through
{the disc of}
M33 is
{about}
twice that through the Milky Way
{(where the observer is located not far from the midplane),
and we adopt $L\approx 1\kpc/\cos(i)\approx 2\kpc$} \citep{2008A&A...490.1005T} for this
galaxy. Then the value of $\delta_I$ in M33, reduced to {a standard value of} $L=1\kpc$ for
compatibility with the Milky Way data, is further $\sqrt2$ times larger:
\[
\delta_I\simeq0.3.
\]

\subsection{Comparison with earlier results and summary}
{To date, analysis of synchrotron observations of the Milky Way has focused}  primarily on the spectrum
of the fluctuations while their magnitude has attracted surprisingly little
attention. \citet{MS57} observed fluctuations of the Galactic radio background
near the Galactic south pole at $\lambda=3.5\m$, with the resolution of
$50\degr$, to obtain $\sigma_I=3.3\times10^{-26}\,\mathrm{W\,m^{-2}}\Hz^{-1}$
per beam, which corresponds to $\delta_I=0.12$ \citep[see also][]{G59}.

\citet{DS87} used observations at $102.5\MHz$ ($\lambda2.92\m$) near the North
Galactic Pole, where the Galactic radio emission is minimum, to determine the
synchrotron autocorrelation function and its anisotropy arising from the
large-scale magnetic field. They obtain $\delta_I\approx 0.07$ but note that
this estimate should be doubled if the isotropic extragalactic background (half
the total flux) is to be subtracted, to yield $\delta_I\simeq0.14$.
\citet{Betal91a} argue that only 17\% of the total flux is of extragalactic
origin, and then $\delta_I\simeq0.08$ at $102.5\MHz$.

The autocorrelation function of the brightness temperature fluctuations of the
Galactic radio background was determined also by \citet{Betal91a,Betal91b}
who used observations at 408 and 1420\,MHz, smoothed to a resolution of about
$\mathrm{FWHM}=5\degr$. They observed a `quiet' region with reduced
fluctuations, $30\degr<\mathrm{DEC}<50\degr$, $180\degr<\mathrm{RA}<250\degr$,
identified by \citet{B67} as an interarm region, since they were interested in
the cosmic microwave background fluctuations. For the Galactic synchrotron
radiation, which dominates at these frequencies, they obtain
$\delta_I\approx0.05$ at 408\,MHz and 0.08 at 1420\,MHz.

These estimates are somewhat lower than those obtained above.
The value of $\delta_I$ obtained by \citeauthor{Betal91a} can be lower due to
their selection of a region with weaker synchrotron intensity fluctuations.

The relative fluctuations in radio intensity are remarkably similar in all the
Milky Way maps and in all fields in M33 considered, with
\[
\delta_I\approx0.1\mbox{--}0.3,
\]
when reduced to the common pathlength $L=1\kpc$ (see Section~\ref{scale} for further discussion).
The lower values are more plausible.
We believe that these
estimates are not significantly affected by either large-scale trends
in the radio intensity or by discrete radio sources or by thermal emission.
Even if these effects still contribute to our estimate, it provides an
\textit{upper\/} limit on the fluctuations in synchrotron intensity arising
in the interstellar medium of the Milky Way and M33.

\section{Statistical analysis of synchrotron intensity fluctuations}\label{SIF}

{In order to interpret the results of our analysis of observations in Section~\ref{data}, we use a model of partially ordered, random distributions of magnetic fields and cosmic rays, assuming various degrees of correlation (or anticorrelation) between them. We calculate the relative magnitude of synchrotron intensity fluctuations $\delta_I$ analytically and numerically and compare the results with the observational constraints described above in order to establish to estimate the degree of correlation or anticorrelation compatible with observations.}

{In this Section, first we relate fluctuations in synchrotron itensity to the underlying synchrotron emissivity, then we describe a model for defining distributions of magnetic field and cosmic rays with controlled cross-correlation. These magnetic field and cosmic ray distributions allow us to calculate the model synchrotron emissivity, using results derived in Appendix~\ref{app}, and hence the model synchrotron intensity.}

\subsection{{Relative fluctuations of synchrotron intensity}}\label{RFSI}
{Here we}
estimate the synchrotron intensity $I_0=\bra{I}_S$
{averaged over an area $S$}
in the plane of the sky and the
{corresponding}
standard deviation,
$\sigma_I=\left(\bra{I^2}_S-\bra{I}_S^2\right)^{1/2}$, as a function of the synchrotron emissivity.
{Angular brackets are used to denote various spatial averages
(over an area $S$, volume $V$ or path length $L$, as indicated by the corresponding
subscript), whereas overbar is used for
statistical (ensemble) averages. The former arise naturally from observations and numerical
models, whilst analytical calculations usually provide the latter. We assume that the two types of averaging procedure
lead to identical results} \citep[unlike ][]{GSFSM13}, although we distinguish them formally for the sake of clarity.

Assuming that the synchrotron spectral index is equal to $-1$
{(this simplifies analytical calculations significantly, without
noticeable effect on the results -- see Section~\ref{Disc} below)},
the intensity
of synchrotron emission at a given position in the sky is given by
\begin{equation}\label{I}
    I=\int_L \varepsilon\,\dd s \propto\int_L n\cra B_\perp^2 \,\dd s ,
\end{equation}
where $\varepsilon$ is the synchrotron emissivity, $n\cra$ is the number
density of cosmic ray electrons, $B_\perp$ is magnetic field in the plane of
the sky and integration is carried along the line of sight $\vec{s}$ over the
path length $L$.

{In a random magnetic field and cosmic-ray
distribution, the synchrotron emissivity $I$ is also a random variable.
We can rewrite Eq.~(\ref{I}) in terms of the path-length average as
\begin{equation}\label{Iapp}
    I = L \bra{\varepsilon}_L\,,
\end{equation}
where $\bra{\varepsilon}_L=L^{-1}\int_L \varepsilon\,\dd s$ is the average
synchrotron emissivity along the path length.
We neglect a dimensional factor in Eq.~\eqref{Iapp} and other similar equations; it is
unimportant as we always consider relative fluctuations where it cancels.
The average synchrotron intensity from the area $S$ in the sky plane is then
related to the synchrotron emissivity averaged over the volume of a depth $L$
(the extent of the radio source along the line of sight) and cross-section $S$:
\begin{equation}\label{I0e}
I_0=\bra{I}_S=L \bra{\varepsilon}_V=2N l_\varepsilon \aver{\varepsilon}\,,
\end{equation}
where $l_\varepsilon$ is the correlation length of the synchrotron emissivity,
$N=L/(2l_\varepsilon)$ $(\gg1)$ is the number of correlation cells of $\varepsilon$
along the path length $L$,
and the volume average has been identified with the statistical average to obtain the last equality,
\begin{equation}\label{Vst}
\bra{\varepsilon}_V=\aver{\varepsilon}\,.
\end{equation}
If $S$ is sufficiently large, such an identification applies to the area average as well,
but the linear resolution of observations often approaches the size of a turbulent cell
in the source; in this case, Eq.~\eqref{Vst} is more appropriate.
}

{Fluctuations in $I$ arise from variations of both $\varepsilon$ and $L$ between different
lines of sight. Neglecting the latter, the standard deviation of $I$ follows as
\begin{equation}\label{sigmaIe}
\sigma_I=N^{-1/2}L\sigma_\varepsilon=\sigma_\varepsilon(2l_\varepsilon L)^{1/2}\,,
\qquad N\gg1\,;
\end{equation}
this quantity characterizes the scatter in the synchrotron intensity
at different positions across the radio source separated by more than $l_\varepsilon$
to make them statistically independent.
}

\subsection{Cosmic ray distribution partially correlated with magnetic field}\label{Cray}
{In this section we introduce a distribution}
of cosmic rays which has a prescribed cross-correlation coefficient with a given
magnetic energy density.
{Magnetic field is represented as the sum of the mean and random parts,
$\vec{B}_0$ and  $\vec{b}$, respectively:}
\[
\vec{B}=\vec{B}_0+\vec{b},
\]
{with $\aver{\vec{b}}=0$, $\aver{\vec{B}}=\vec{B}_0$ and
$\aver{B^2}=B_{0\perp}^2+B_{0\parallel}^2+\sigma_b^2$, where $B_{0\perp}$ and
$B_{0\parallel}$ are the mean field components in the plane of the sky and
parallel to the line of sight, respectively.}
Each Cartesian vectorial component of the random magnetic field $\vec{b}$
is assumed to be a Gaussian random variable with zero mean value, $\aver{b}_i=0$, and
{to avoid unnecessary complicated calculations,}
the random magnetic field is assumed to be isotropic,
\begin{equation}\label{biso}
\aver{b_i^2}=\tfrac{1}{3}\aver{b^2}=\tfrac{1}{3}\sigma_b^2,
\end{equation}
where $i=x,y,z$ and overbar denotes ensemble averaging.

{The number density of cosmic-ray electrons is similarly represented as the
sum of a mean $n_0\equiv\aver{n}\cra$ (slowly varying in space) and random $n'$
parts,}
\begin{equation}\label{ncra0}
n\cra=n_0+n'.
\end{equation}
The cross-correlation coefficient $c$ of $n\cra$ and $B^2$ is defined as
\begin{equation}\label{corr}
c(B^2,n\cra)=
\frac{\aver{ n\cra B^2} -\aver{n}\cra\aver{B^2}}{\sigma_{n} \sigma_{B^2}}.
\end{equation}

{
To implement local equipartition between cosmic rays and magnetic field,
which corresponds to $c=1$, we use a distribution of cosmic rays
identical to that of the random part of the magnetic energy density,
$n'=\alpha(B^2-\aver{B^2})$,
where $\alpha$ is a coefficient that allows us to control independently the
magnitude of the fluctuations in $n\cra$.

To obtain partially correlated distributions of $n\cra$ and $B^2$, we introduce
an auxiliary positive-definite, scalar random field $F$, uncorrelated with the magnetic
energy density:
\begin{equation}\label{cB2F}
c(B^2,F)=0.
\end{equation}
Our specific choice of $\vec{B}$ and $F$ is discussed below.

Now, we represent the random part of the cosmic-ray number density as
\begin{equation}\label{ncra}
n'=\alpha (B^2-\aver{B^2}) + \beta(F-\aver{F}),
\end{equation}
where the coefficients $\alpha$ and $\beta$ are chosen to obtain
\begin{equation}\label{cC}
c(B^2,n\cra)=C,
\end{equation}
with $C$ the desired value of the cross-correlation coefficient.
The first term in Eq.~\eqref{ncra} is responsible for the part of the cosmic-ray
distribution fully correlated with magnetic field energy density,
whereas the second term reduces the cross-correlation to the desired level.
In particular,  Eq.~\eqref{ncra} ensures that $c(B^2,n\cra)=1$ for $\alpha=1$,
$\beta=0$ (perfect correlation),
and $c(B^2,n\cra)=0$ for $\alpha=0$ (uncorrelated distributions).

Let us find $\alpha$ and $\beta$ from Eq.~\eqref{cC}.
It follows from Eqs~(\ref{ncra0}) and (\ref{ncra}) that
\[
\aver{n\cra B^2}=
\alpha \sigma^2_{B^2}+n_0 \aver{B^2}
\]
given that Eq.~\eqref{cB2F} implies
\[
\aver{B^2 F}= \aver{B^2}\, \aver{F}.
\]
For any uncorrelated random variables $X$ and $Y$, the variance of their sum
$Z=X+Y$ is given by $\sigma_Z^2=\sigma_X^2+\sigma_Y^2$. Hence, Eq.~\eqref{ncra}
implies
\begin{equation}\label{sigman}
\sigma^2_{n}=\alpha^2 \sigma^2_{B^2} + \beta^2 \sigma^2_{F}.
\end{equation}
Then Eq.~\eqref{cC} yields
\begin{equation}\label{aC}
\frac{\alpha}{\sqrt{\alpha^2\sigma^2_{B^2}+\beta^2\sigma^2_{F}}}=C,
\end{equation}
where $\sigma_{B^2}=\aver{B^4}-(\aver{B^2})^2$ and $\sigma_F^2=\aver{F^2}-\aver{F}^2$.
Using Eq.~\eqref{sigman} to
eliminate $\beta$ in Eq.~\eqref{aC}, we obtain
\[
\alpha=\frac{\sigma_n}{\sigma_{B^2}}C
   \quad\mbox{and}
   \quad
\beta=\frac{\sigma_n}{\sigma_{F}}\sqrt{1-C^2 }.
\]
Equation~\eqref{ncra} then reduces to
\begin{equation}\label{ncrafull}
n\cra=n_0+\sigma_n\left( C \frac{B^2-\aver{B^2}}{\sigma_{B^2}}+\sqrt{1-C^2}\frac{F-\aver{F}}{\sigma_{F}} \right),
\end{equation}
where the standard deviation of the cosmic-ray number density $\sigma_n$ is an independent parameter
that we are free to vary.

Our specific model of magnetic field is described in Section~\ref{MFCRM}. However, for the analytical
calculations of $\delta_I$ presented in Section~\ref{SIFPC}, it is sufficient to know $B_0$ and $\sigma_B$.
There is no need to specify $F$ in any more detail as long as $F$ and $B^2$ (more precisely, $B_\perp^2$)
are uncorrelated. The more
detailed model of Section~\ref{MFCRM} is only required for numerical calculations presented
below to verify and refine the analytical results.

\subsection{Synchrotron intensity fluctuations with partially correlated $n\cra$ and $B^2$}\label{SIFPC}
Details of the calculations of $\aver{\varepsilon}$ and $\sigma_\varepsilon$, and then, of the mean value
and standard deviation of the synchrotron intensity
using Eqs~\eqref{I0e} and \eqref{sigmaIe} together with Eq.~\eqref{ncrafull}
can be found in Appendix~\ref{app}, with $I_0$ given by Eq.~(\ref{I0}) and $\sigma_I$ by Eq.~(\ref{sigmaI}).
Here we consider the key special cases.

The model developed here allows us to express $\delta_I$ in terms of the following dimensionless parameters:
the number of correlation cells along the line of sight $N$
(assuming perfect angular resolution; a finite beam size can be allowed for using
additional averaging across the beam as in
the end of Section~\ref{ssM33}), the cross-correlation coefficient between
cosmic rays and magnetic field $C$, the relative magnitude of the magnetic field fluctuations $\delta_b$,
and the relative magnitude of fluctuations in cosmic ray number density,}
\[
\delta_n=\frac{\sigma_n}{n_0}.
\]

\begin{figure}
\includegraphics[width=0.4\textwidth]{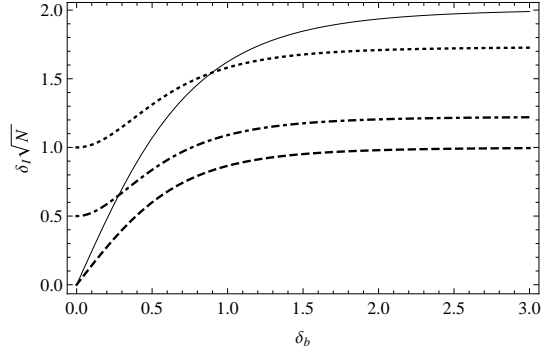}
\caption{\label{I1b} The relative fluctuations in synchrotron intensity
$\delta_I$ versus the relative magnitude of magnetic field fluctuations
$\delta_b$
{obtained analytically in four special cases.
Solid line, Eq.~\eqref{dIeq}, is for complete correlation of cosmic rays and
magnetic fields, $C=1$;
dashed line is from Eq.~\eqref{homogncra}, i.e., $n\cra=\mbox{const}$;
dotted and dash-dotted lines were obtained
from Eq.~\eqref{bcrauncorr} for
uncorrelated fluctuations in cosmic rays and magnetic field, $C=0$, with
$\delta_n=1$ and $\delta_n=0.5$, respectively.} We note that the curves rapidly
approach the horizontal asymptote,
{so that} the approximation $\delta_b\to\infty$
is reasonably accurate for $\delta_b>1.5$--2
{as typically found in spiral galaxies}.}
\end{figure}

In the case of detailed (local, or pointwise) equipartition between cosmic rays and magnetic fields,
$n\cra\propto B^2$, {$C=1$ and $\delta_n=\sigma_{B^2}/B^2$ (Eq.~(\ref{eq:App:sigB2}))} so the relative fluctuations in the synchrotron intensity follow as
\begin{equation}\label{dIeq}
\delta_I=\frac{\sigma_I}{I_0}=
\frac{\delta_b(54+295\delta_b^2+404\delta_b^4+101\delta_b^6)^{1/2}}{N^{1/2}\left(3+10\delta_b^2+5\delta_b^4\right)}.
\end{equation}
We recall that $\delta_b=\sigma_b/B_0$
{and note that all such analytical
results are valid only for $N\gg1$.}
As $\delta_b$ increases, $\delta_I$ rapidly
approaches the asymptotic value independent of $\delta_b$ (see Fig.~\ref{I1b}):
\[
\delta_I\simeq 2N^{-1/2} \quad \mbox{for}\quad \delta_b^2\gg1.
\]

It is useful to note similar expressions for $\delta_I$ obtained under
different assumptions about the correlation between cosmic rays and magnetic
field. If cosmic ray fluctuations are
{uncorrelated with those} in
magnetic field, Eqs.~(\ref{I0}) and (\ref{sigmaI}) yield
for $C=0$
\begin{equation}\label{bcrauncorr}
\delta_I=
\frac{[\delta_n^2+\delta_b^2(2+\delta_b^2)(1+2\delta_n^2)]^{1/2}}{N^{1/2}\left(1+\delta_b^2\right)}.
\end{equation}
In particular, for $\delta_n\to0$ we obtain an asymptotic form for a
homogeneous distribution of cosmic rays:
\begin{equation}\label{homogncra}
\delta_I=
\frac{\delta_b(2+\delta_b^2)^{1/2}}{N^{1/2}\left(1+\delta_b^2\right)}.
\end{equation}
Thus, $\delta_I\simeq N^{-1/2}$ for $\delta_b^2\gg1$ and $n\cra=\mbox{const}$.
Figure~\ref{I1b} shows the dependence of $\delta_I$ on $\delta_b$ from
Eqs~\eqref{dIeq}, (\ref{bcrauncorr}) and  (\ref{homogncra}).

{It is convenient to summarize these results by providing the corresponding
values of a quantity independent of the number of correlation cells along the
path length, $N^{1/2}\delta_I$, as obtained from
Eqs~\eqref{dIeq}, (\ref{bcrauncorr}) and  (\ref{homogncra}) for $\delta_b\gg1$,
which is applicable to $\delta_b\simeq3$ (Fig.~\ref{I1b}).}
The relative magnitude
of synchrotron intensity fluctuations expected under detailed equipartition follows from
Eq.~(\ref{dIeq}) as
\begin{equation}\label{complete}
N^{1/2}\delta_I\approx2.0
\quad\mbox{(local equipartition)}.
\end{equation}
Equation~(\ref{bcrauncorr}) yields, for $\delta_n=0.5$,
\[
N^{1/2}\delta_I\approx1.2
\quad\mbox{(uncorrelated fluctuations)},
\]
and Eq.~(\ref{homogncra}) leads to
\[
N^{1/2}\delta_I\approx1.0
\quad (n\cra=\mbox{const}).
\]
As might be expected, detailed equipartition between cosmic rays and magnetic
fields leads to the strongest synchrotron intensity fluctuations for a given $\delta_b$
and $N$. {For illustration, with the correlation length of the synchrotron intensity fluctuations
 $\corr_\varepsilon=50\p$ and the path length $L=1\kpc$, we
obtain $N=L/2\corr_\varepsilon\simeq 10$ for a beam narrower than the
 size of the correlation cell.} We note that the dependence of
$\sqrt{N}\delta_I$ on $\delta_b$ is quite weak as long as  $\delta_b^2\ga3$ {which is usually the case for spiral galaxies (see Section ~\ref{MFIMF}).
The difference in the level of synchrotron fluctuations
in these limiting cases is strong enough to be observable under certain
conditions clarified in Section~\ref{SR}.}

\section{A model of a partially ordered magnetic field}\label{MFCRM}
{ To verify, strengthen and refine the analytical calculations presented above,
we implement numerically the model of magnetic field and cosmic rays
described in Section~\ref{Cray}. For this purpose, we introduce in this section a
multi-scale magnetic field with prescribed spectral properties and the corresponding
cosmic-ray distribution using Eq.~\eqref{ncrafull}. The phases and directions of
individual modes in the magnetic field spectrum can be chosen at random without affecting
the magnetic field correlation scale,
the value of $\delta_B$ and the energy spectrum. We use this freedom to generate
a large number of statistically independent realizations of the magnetic field and cosmic-ray
distributions to compute the resulting values of $\delta_I$ and compare them with the
analytical results.
}

To prescribe a quasi-random magnetic field $\vec{b}$ with vanishing mean value
in a periodic box, we use a Fourier expansion in modes with randomly chosen
directions of wave vectors $\vec{k}$, but with amplitudes adjusted to
reproduce any desired energy spectrum:
\begin{equation}\label{ft}
\vec{b}(\vec{x})=\frac{1}{(2\pi)^{3/2}}\int \vec{\hat{b}}(\vec{k})
e^{\mathrm{i} \vec{k}\cdot\vec{x}}\, \dd^3\vec{k},
\end{equation}
where $ \vec{\hat{b}}$ is the Fourier transform of $\vec{b}$;
the physical field is represented by the real part of this complex vector. The
corresponding magnetic energy spectrum is given by
\begin{equation}\label{spec}
M(k)=\int_{|\vec{k}'|=k} |\vec{\hat{b}}(\vec{k}')|^2\, \dd^3\vec{k}',
\end{equation}
where the integral is taken over the spherical surface of
a radius $k$ in $k$-space. In the isotropic case, $M(k)=4\pi k^2
|\vec{\hat{b}}(k)|^2$. In order to ensure periodicity within a computational
box of size $L$, as required for the discrete Fourier transformation, the
components of the wave vectors are restricted to be integer multiples of
$2\pi/L$.

A solenoidal vector field $\vec{b}$, i.e., that having
$\vec{k}\cdot\vec{\hat{b}}(\vec{k})=0$, is specified by
\[
\vec{\hat{b}}(\vec{k})=\frac{\vec{k}\times\vec{X}}{|\vec{k}\times\vec{X}|}k^{-1}\sqrt{M(k)},
\]
where $\vec{X}$ is a complex vector chosen at random, to ensure that the Fourier
modes have random phases.
We consider a magnetic energy spectrum represented by two power-law ranges,
\begin{equation}\label{Mss}
M(k)=M_0
\left\{
\begin{array}{ll}
(k/k_0)^{s_0}    &\mbox{for } k<k_0,\\
(k/k_0)^{-s_1}    &\mbox{for } k\geq k_0,
\end{array}
\right.
\end{equation}
with $s_0>0$ and $s_1>0$, where $k_0$ is the energy-range wavenumber. We use
$s_1=5/3$ as in Kolmogorov's spectrum and $s_0=2$
\citep[see][]{2001PhRvE..64e6405C}. The standard deviation of the magnetic
field is given by
\begin{equation}\label{sigmabs}
\sigma_b^2=\int_0^\infty M(k)\,\dd k=M_0 k_0\frac{s_0+s_1}{(s_0+1)(s_1-1)}
\end{equation}
for $s_1>1$.
The correlation length $\corr_b$ of the resulting magnetic field
(analogous to the radius of a correlation cell) differs from its dominant
half-wavelength $\tfrac12\lambda_0=\pi/k_0$ for any finite values of $s_0$ and
$s_1$ {\citep{Monin:1975}}:
\begin{align}
\corr_{b}&
=\frac{\pi}{2}\,\frac{\int_0^\infty k^{-1}M(k)\,dk}{\int_0^\infty M(k)\,\dd k}
= \frac{\pi}{2k_0} \left(1+\frac{1}{s_0}\right)\left(1-\frac{1}{s_1}\right)
                                                                \nonumber\\
&=\frac{3\pi}{10k_0},\label{Bcorr}
\end{align}
where the last equality follows for $s_0=2$ and $s_1=5/3$. {For the Milky Way, suitable values are $l_b=50\p$ and $\sigma_b\simeq 5$--$10\mkG$ (see Sections \ref{MFIMF} and \ref{Disc}).}

The resulting solenoidal vector field is then added to a uniform component
$\vec{B}_0$ to produce a partially ordered magnetic field with controlled
fluctuation level $\delta_b$ and energy spectrum $M(k)$. This approach
has been used to generate synthetic polarization maps of the turbulent ISM
by
\citet{2008A&A...480...45S,2010JETPL..90..637V,2011AN....332..524A,2011arXiv1109.4062M}.
Similar constructions were used by \citet{GJ99} and \citet{CLP02} in their
modelling of cosmic ray propagation in random magnetic fields, by \citet{MV99}
for modelling turbulent flows, and by \citet{WBS07} to study dynamo action in
chaotic flows.

\begin{figure}
 \includegraphics[width=0.4\textwidth]{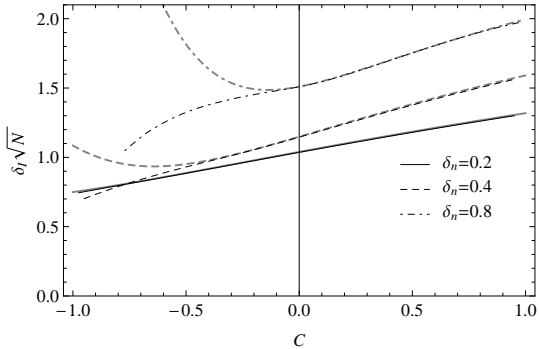}
 \caption{\label{I1a} The relative fluctuations in synchrotron intensity
$\delta_I$ as obtained from analytical formulae Eqs.~(\ref{I0} and \ref{sigmaI})
(thicker curves) and from numerical calculations with the condition $n\cra>0$
enforced (the corresponding thinner curves), for:
$\delta_n=0.2$ (solid),
$\delta_n=0.4$ (dashed) and
$\delta_n=0.8$ (dash-dotted).
The analytic formulae become inapplicable for $C<0$ and $\delta_n\simeq1$ (see
the text). The magnetic field is purely random, $B_0=0$ and $N=10$.}
\end{figure}

We will now verify, by direct calculation, that $C(B^2,n\cra)\approx C$.
The reason for the approximate equality is first explained.

A shortcoming of the { analytical} cosmic ray model defined by Eqs.~(\ref{ncra0}) and (\ref{ncra}) is that $n\cra$ can be negative at some
positions (especially when $C<0$ and hence $\alpha<0$) because, at some
positions and in some realizations, $B^2$ can be arbitrarily large (as a
Gaussian random variable squared). This deficiency could be corrected by
selecting a more realistic probability distribution for $\vec{b}$ (e.g., a
truncated Gaussian) but we do not feel that this would lead to any additional
insight. In the numerical calculations described below, we truncate the negative
values of $n\cra$ by replacing them with zero. This, however, makes it impossible
to achieve exact anticorrelation between cosmic rays and magnetic fields, so
that $C(B^2,n\cra)>-1$.

In analytical calculations, we restrict ourselves to the cases with
$\delta_n<1$ to reduce the extent of the problem (even if not to
resolve it completely). For example, Eqs.~(\ref{I0}) and (\ref{sigmaI}) yield
for $\delta_b\to\infty$ {(note that $\delta_I$ is constant with respect to $\delta_b$ for $\delta_b\gtrsim 3$ (Fig.~\ref{I1b}))}
\begin{equation}\label{delIB0}
\delta_I=
\frac{\left[9+6\delta_n\left(11C^2\delta_n+3\sqrt{6}C+3\delta_n\right)\right]^{1/2}}{N^{1/2}(\sqrt{6} C \delta_n +3)}.
\end{equation}
This dependence of $\delta_I$ on the cross-correlation coefficient $C$ is
shown with thicker curves in Fig.~\ref{I1a}
for various values of the relative
fluctuations in cosmic rays, $\delta_n$. Thinner curves show similar results
obtained from a numerical calculation where $n\cra>0$ is enforced. It is clear
from Fig.~\ref{I1a} that these analytical results are accurate for $C>0$.

However, for $C<0$, {analytic results}
are useful only  if the fluctuations in the cosmic ray
number density are relatively weak: $C\ga-0.1$ for $\delta_n<0.8$, $C\ga-0.5$
for $\delta_n<0.4$, and  almost any value of $C$ for $\delta_n<0.2$. We only use these
analytical results for illustrative purposes, whereas all our conclusions are
based on numerical results where $n\cra>0$ at all positions. {Nevertheless, the analytic results presented here and in Appendix~\ref{app}, however unwieldly, are simpler to use than constructing a numerical model and are accurate for $\delta_n$ small enough (say, $\delta_n\lesssim0.4$) and $C$ large enough (say, $C\gtrsim-0.5$) --- see Fig. ~\ref{I1a}.}

\section{Results} \label{SR}
\subsection{Synthetic radio maps}\label{SyRaMa}
Each component of the magnetic field described by Eq.~(\ref{ft}) is
the sum of a large number of independent contributions from different
wave numbers. By virtue of the central limit theorem, each component of the
resulting magnetic field is well approximated by a Gaussian random variable.
Then the mean synchrotron intensity and its standard deviation over $N$
correlation cells can be expressed, using Eq.~(\ref{I}), in terms of $B_0$,
$\sigma_b$, $\delta_n$ and $C$. Explicit analytic expressions for $I_0$
and $\sigma_I$ can be found in Appendix~\ref{app}, and we illustrate these
results in Fig.~\ref{I1a}. As might be expected, the relative level of the
synchrotron intensity fluctuations increases with the cross-correlation
coefficient between $B^2$ and $n\cra$.

Since analytical results are of limited relevance for $C<0$, we performed
numerical calculations of the synchrotron intensity where the cosmic-ray number density
is truncated to be non-negative (i.e. $n\cra=0$ wherever the model defined by Eq.~(\ref{ncra0}) returns
a negative value). The model has four free parameters:
\begin{enumerate}
\item[(i)] the relative level of magnetic field fluctuations  $\delta_b=\sigma_b/B_0$,
\item[(ii)] the relative level of cosmic-ray number density fluctuations
$\delta_n=\sigma_n/n_0$,
\item[(iii)] the cross-correlation coefficient between magnetic field and
cosmic rays $C$, and
\item[(iv)] the {dominant energy wave number} of the turbulent magnetic field $k_0$.
\end{enumerate}
We do not vary the spectral index of magnetic field and cosmic rays as these
parameters are of secondary importance {in this context}.

The value of $k_0$ controls the correlation lengths of magnetic field (Eq.~(\ref{Bcorr})),
cosmic rays and synchrotron emissivity, and hence the number of the
correlation cells of synchrotron intensity fluctuations in the telescope beam $N$, which
in turn affects the magnitude of synchrotron intensity fluctuation as
$\delta_I\propto N^{-1/2}$. Since $N$ can vary widely between different lines
of sight in the Milky Way and between galaxies with different inclination
angles, we present our results in terms of $N^{1/2}\delta_I$
for both the observations and the model.

\begin{figure}
 \includegraphics[width=0.45\textwidth]{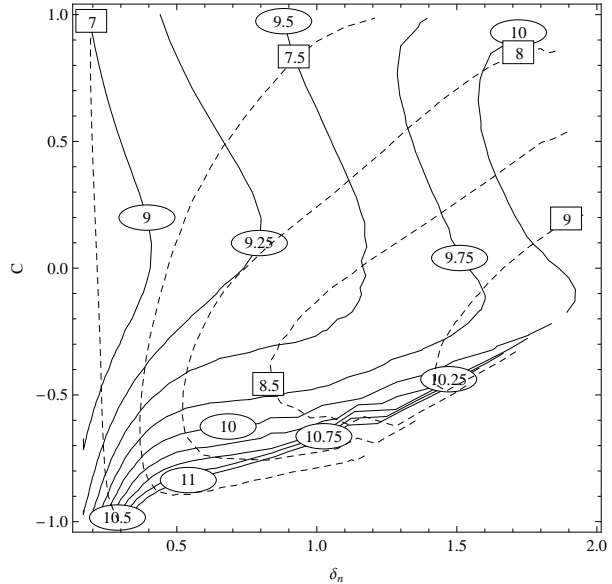}
 \caption{\label{figN} The isocontours of the numerical factor $G$ in
Eq.~(\ref{prefacG}) for $\sigma_b=1$ (dashed lines and labels in
squares) and $\sigma_b\to\infty$ (solid lines and labels in ovals) shown in
the $(\delta_n,C)$-plane.}
\end{figure}

\subsection{A relation between the correlation lengths of the synchrotron emissivity and magnetic field}\label{scale}
In the case of an infinitely narrow beam, the
number of synchrotron correlation cells traversed by the emission
is just the ratio $N=L/(2 \corr_\varepsilon)$, where $L$ is the
path length through the synchrotron source and $\corr_\varepsilon$ is
the correlation length of the fluctuations in the synchrotron emissivity. For
a finite beam width $W$, this is the number of correlation cells within the
beam cylinder, $N\simeq(3/16)LW^2/l_\varepsilon^3$,
assuming a circular beam and spherical correlation cells.
Unlike the correlation lengths of physical parameters such as the magnetic
field, velocity or density fluctuations, the correlation length of the
intensity (or emissivity) variations cannot be deduced independently
(e.g., from the nature of the turbulence driving), but has
to be calculated from the statistical parameters of the physical variables or
from observations. {Here, we shall derive an expression for $\corr_\varepsilon$ in terms of $\corr_b$, which will allow us also to estimate $N$.}

To illustrate the difficulties arising, consider the autocorrelation function
of $b_\perp^2$ as an example. If $V(x)$ is a stationary Gaussian random
function, with vanishing mean value and the autocorrelation function
$K_v(r)=\aver{V(x)V(x+r)}$, the {autocorrelation function of $V^2(x)$ is
given by $K_{v^2}(x)=2[K_v(r)]^2$} \citep[see e.g.\ \S13 in][]{S66}. Assuming that
each {Cartesian vector} component $b_i$ of the random magnetic field $\vec{b}$ is a Gaussian
random variable, with the autocorrelation function $K_{b_i}$, we have
$K_{{b_i}^2}=2K_{b_i}^2$. Assuming statistical isotropy of $\vec{b}$,
$K_{b_x}(r)=K_{b_y}(r)$, and neglecting any cross-correlations between $b_x$
and $b_y$, we obtain the autocorrelation function of $b_\perp^2$:
\[
K_{b_\perp^2}(r)=4 K_{b_i}^2(r).
\]
The relation between the
correlation scales of $b_i$ and $b_\perp^2$, denoted $\corr_{b_i}$ and
$\corr_{b_\perp^2}$, respectively, depends on the form of the autocorrelation
function of magnetic field: for $K_{b_i}=\tfrac13\sigma_b^2\exp(-r/\corr_b)$,
we have $\corr_{b_\perp^2}=\corr_{b_i}/2$. However, for
$K_{b_i}=\tfrac13\sigma_b^2\exp(-\pi r^2/\corr_b^2)$ we have
$\corr_{b_\perp^2}=\corr_{b_i}/\sqrt2$.
There is no universal relation between
the correlation scales of even these simply connected variables. Such a
relation should be established in each specific case from the statistical
properties of each physical component of the system.

{Since the power spectrum is a Fourier transform of the correlation function, these arguments also apply to the power spectra of $b_i$ and $\epsilon$.}

We calculate the correlation length $\corr_\varepsilon$ of the synchrotron
emissivity $\varepsilon\propto n\cra B_\perp^2$  in the synthetic radio maps
from its autocorrelation function $K_\varepsilon(r)$, for various values of
the cross-correlation coefficients $C$, $B_0$ and $n_0$:
\begin{equation}\label{lcorr}
\corr_\varepsilon
=\sigma_\varepsilon^{-2} \left(
\int_{0}^{L} K_\varepsilon(r)\, \dd r
        - \bra{\varepsilon}^2
\right),
\end{equation}
where $L$ is the path length  and
$\sigma_\varepsilon$ is the standard deviation of the synchrotron emissivity (assuming $L\gg \corr_\varepsilon$ to minimize the impact of statistical fluctuations).

The resulting dependence of $\corr_\varepsilon$ on the {correlation length} of
the magnetic field $\corr_b$, obtained in Eq.~(\ref{Bcorr}) {for the spectrum given by Eq.~(\ref{Mss})}, can
be approximated as
\begin{equation}
\corr_\varepsilon/L\approx G^{-1}(4\corr_b/L)^{0.65},
\label{prefacG}
\end{equation}
where the numerical factor $G$ depends on the model parameters and the
cross-correlation coefficient $C$. The contours of $G$ in the
$(\delta_n,C)$-plane are shown in Fig.~\ref{figN}; $G=9$--10 are
representative values for $\delta_b>2$--3,
{independent of the exact choice for $\delta_n$ provided $\delta_n\la1$}.
The resulting values of
$N=L/(2\corr_\varepsilon)$ are used below to compare the synchrotron
intensity fluctuations obtained from observations in Section~\ref{data} with the model
of Section~\ref{MFCRM}.

\begin{figure}
 \includegraphics[width=0.4\textwidth]{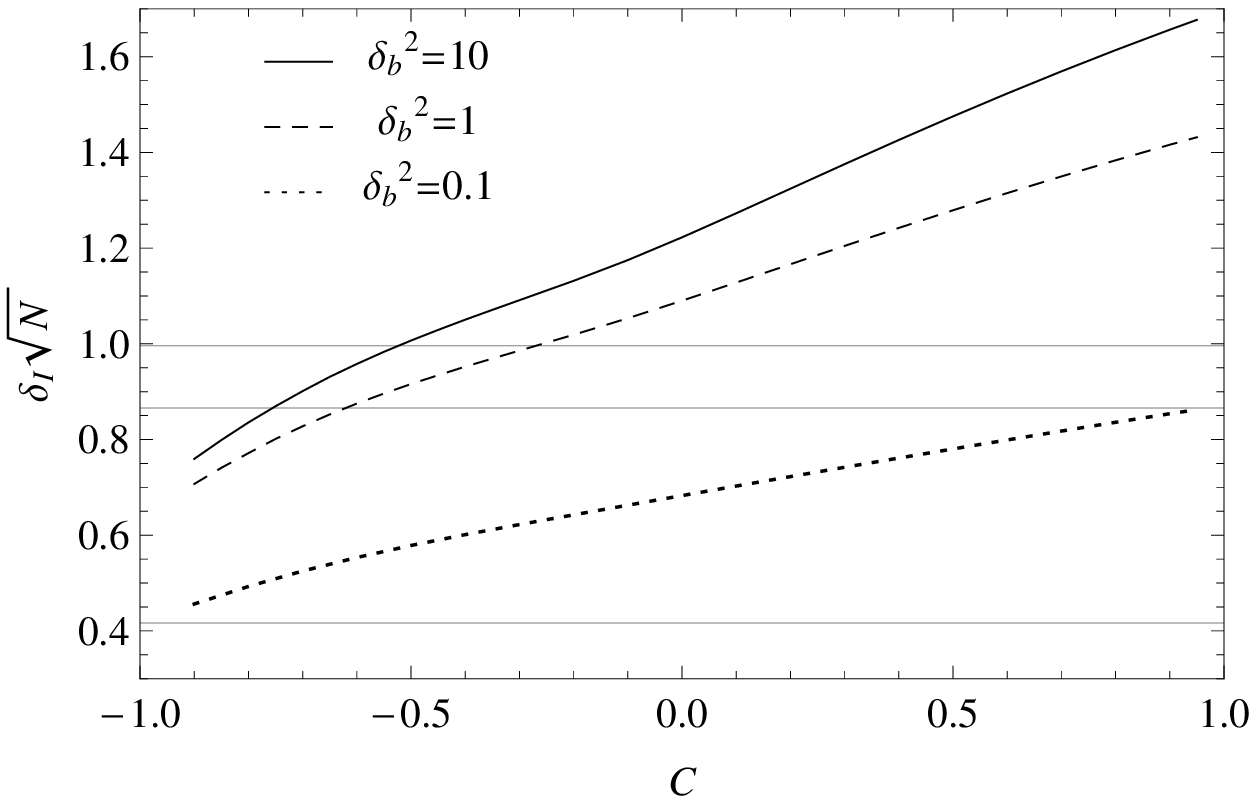}
 \includegraphics[width=0.4\textwidth]{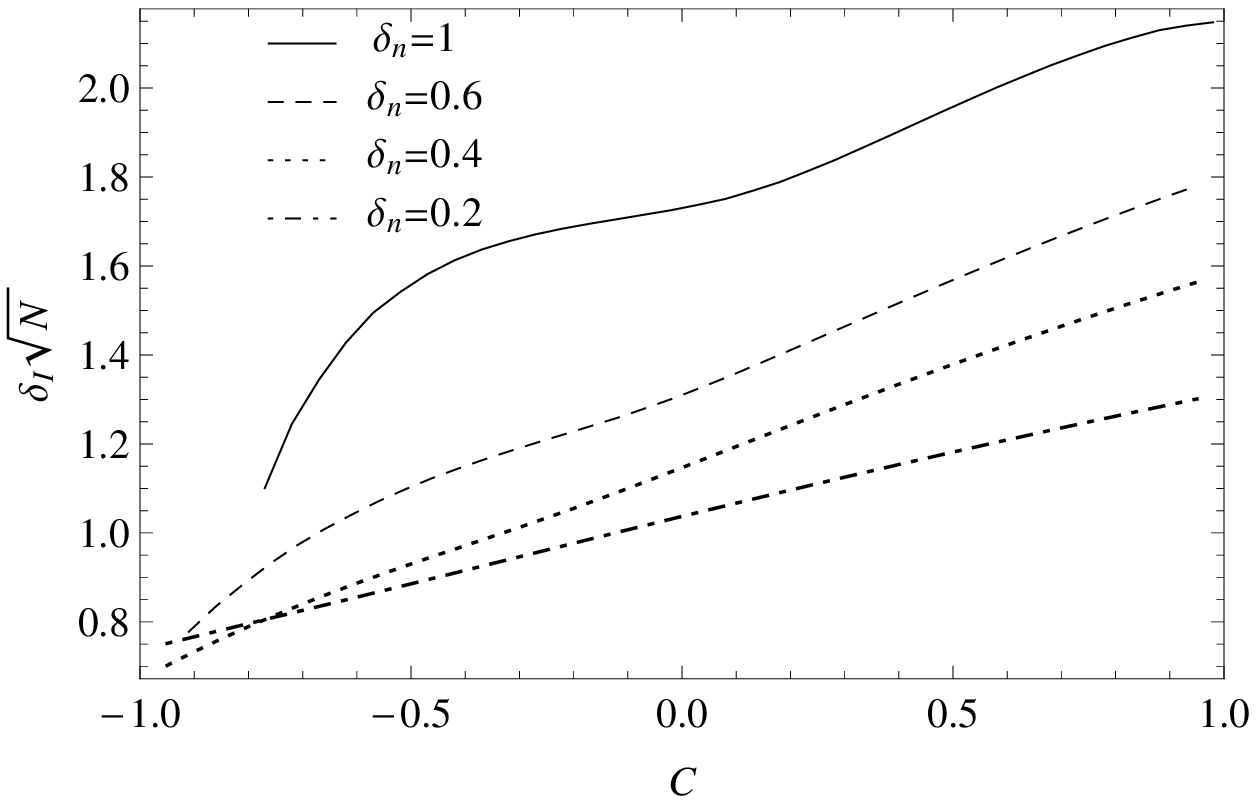}
 \caption{\label{I1} Relative fluctuations of the synchrotron intensity
in synthetic radio maps of Section~\ref{SyRaMa} versus the
cross-correlation coefficient between $B^2$ and $n\cra$ for various choices of
the model parameters. Top panel: a selection of $\delta_b$ values for
fixed $\delta_n=0.5$ (solid: $\delta_b^2=10$; dashed: 1; dotted: 0.1).  Grey
horizontal lines correspond to uniformly distributed cosmic
rays, $\delta_n=0$, for the same values of $\delta_b$ (with $\delta_I$
decreasing with $\delta_b$).  Lower panel: various
values of $\delta_n$  for $\delta_b^2=10$
($\delta_n=1$, solid; 0.6, dashed; 0.4, dotted; 0.2, dash-dotted).}
\end{figure}

\subsection{Correlation between cosmic rays and magnetic fields} \label{subsec:corr}
The relative intensity of synchrotron intensity fluctuations is sensitive to the number
$N$ of correlation cells of synchrotron emissivity within the beam (or along
the line of sight in case of a pencil beam). When comparing the
theoretical model with observations, we adopt $L=1\kpc$ for the pathlength in
the Milky Way, $\corr_{b_i}=50\p$ for the correlation length of magnetic
field, $\delta_b=3$ (the asymptotic limit $\delta_b\gg1$ is
quite accurate in this case), and explore the range $-1\leq C\leq1$ for the
cross-correlation coefficient between cosmic-ray and magnetic fluctuations.
For $\corr_b/L=0.05$ and $G\approx9$ (see Fig.~\ref{figN}), we have
$\corr_\varepsilon\approx40\p$ and $N=L/(2\corr_\varepsilon)\simeq10$;
we also discuss the effect of larger values of $N$.

Figure~\ref{I1} shows the dependence of $\delta_I \sqrt{N}$ on the
cross-correlation coefficient $C$ for various values of the
parameters $\delta_b$ and $\delta_n$. The calculations are based on 100
realizations of $\vec{B}$, so
the statistical errors of the mean values shown are negligible.

As can be seen from the upper panel of Fig.~\ref{I1}, the relative magnitude of
synchrotron intensity fluctuations, $\delta_I\sqrt{N}>0.7$, obtained for $\delta_b\geq1$
and $\delta_n=0.5$, is stronger than what is observed in the Milky Way,
$\delta_I\sqrt{N}=0.3$--0.6 assuming $N=10$. If $N=20$, the conservative observational
estimate $\delta_I=0.1$--0.2 translates into $\delta_I\sqrt{N}=0.4$--0.9,
implying $C\la-0.6$ { for the highest $\delta_I \sqrt{N}$}.
Thus, $\delta_n<0.5$ seems to be justified, unless $N$ is significantly larger
than 10 or, otherwise, $\delta_b<1$ (which is highly implausible).
Since the estimate $\delta_I=0.1$--0.2 has been obtained for high
Galactic latitudes, the path length is unlikely to be much longer than 1\kpc,
and the correlation length of the synchrotron intensity fluctuations can hardly be much
shorter than about 50\,pc.
Thus, excluding the case of simultaneously large $L$ and small
$\corr_\varepsilon$, we conclude that the distribution of cosmic ray electrons
is unlikely to have any significant variations at scales of order 50--100\,pc.

The lower panel in Fig.~\ref{I1}, where a range of values of $\delta_n$ are
used with $\delta_b^2=10$, suggests that any positive correlation between
cosmic rays and magnetic fields is only compatible with the observational
estimate $\delta_I\sqrt{N}=0.3$--0.6 (for $N=10$) if $\delta_n<0.2$. In
fact, {a upper limit $\delta_I\sqrt{N}=0.9$ (for $N=20$)  might be achieved for
$n\cra=\mbox{const}$}. The values of $\delta_I\sqrt{N}$ in this
case are shown with grey horizontal lines in the upper panel: for example,
$\delta_I\sqrt{N}=1$ is compatible with $\delta_b^2=10$. However, the lower
values of synchrotron intensity fluctuations in the Milky Way in the range
$\delta_I\sqrt{N}=0.3$--0.9 for $N=10$--20 {can be compatible with} the
presence of fluctuations in cosmic ray density mildly anticorrelated with
those in magnetic field. It is difficult to be precise here, but
$\delta_n<0.2$ and $C<-0.5$ seems to be an acceptable combination of
parameters.

\section{Propagation of cosmic rays and equipartition with magnetic fields}\label{SDCR}
To illustrate the relation between cosmic rays and magnetic fields,
consider a simple model of cosmic ray propagation near a magnetic flux tube.
The number density of cosmic rays $n\cra$ (or their energy density
$\epsilon\cra$) is assumed to obey the diffusion equation with the source $Q$
and diffusivity $D$ terms depending on the magnetic field \citep{P69,KP83,SL85}.
Consider the steady state of a one-dimensional system with $Q=\mbox{const}$.
The magnetic field is assumed to have a statistically uniform fluctuating
component, $\aver{b^2}=\sigma_b^2=\mbox{const}$, whereas the mean field is
confined to a Gaussian slab of half-thickness $L$ symmetric with respect to
$x=0$: $B_y=B_0\exp[-x^2/(2d^2)]$, $B_x=B_z=0$. The cosmic ray diffusivity
is assumed to depend on the relative strength of magnetic fluctuations,
$D=D_0\sigma_b^2/B_y^2$. The resulting steady-state diffusion equation
\[
\frac{\dd}{\dd x}D\frac{\dd}{\dd x}n\cra + Q=0,
\]
can easily be solved with the boundary conditions
\[
\frac{\dd n\cra}{\dd x}(0)=0,\quad n\cra|_{|x|\to\infty}=0,
\]
to yield
\begin{equation}\label{esol}
n\cra= \frac{QB_0^2d^2}{2D_0\sigma_b^2}e^{-x^2/d^2}.
\end{equation}
The total number (and energy) of cosmic rays remains finite despite the
uniform distribution of their sources, $Q=\mbox{const}$, because $D\to\infty$
as $|x|\to\infty$ in this illustrative model.

This simple solution shows that, near a magnetic flux tube in a statistically
homogeneous random magnetic field, cosmic rays concentrate where the total
magnetic field is stronger because their diffusivity is smaller there. In
this example, the spatial distributions of cosmic rays and magnetic field
are tightly correlated.

Another type of argument relating cosmic ray energy density to parameters of
the interstellar medium was suggested by \citet{PS05}. If both the magnetic
flux and the cosmic ray flux are conserved, $BS=\mathrm{const}$ and $n\cra
US=\mathrm{const}$ (where $B$ is the magnetic field strength and $S$ is the
area within a fluid contour, and $U$ is the cosmic ray streaming velocity),
one obtains $n\cra U/B=\mathrm{const}$, which yields $n\cra\propto n^{1/2}$,
given that $U=V_\mathrm{A}\propto B n^{-1/2}$, with $n$ the gas number density
and $V_\mathrm{A}$ the Alfv\'en speed. Thus, the cosmic ray density is independent of the magnetic field strength, and scales with the thermal gas density.
This result relies on the fact that the streaming velocity of the cosmic rays
is proportional to the Alfv\'en speed. If, instead, $U=V$, with $V$ the gas
speed, we obtain $n\cra\propto B$ from these arguments. No clear scaling of the cosmic rays energy density
$\epsilon\cra$ with the magnetic field was observed in the simulations of
\citet{SBMS06} who use the gas velocity for $U$. There is indication that the average propagation length of CREs depends on the degree of field ordering and hence varies between galaxies \citep{2013arXiv1307.6253T}.

Assumption of a detailed, point-wise (local) equipartition between cosmic
rays and magnetic fields is dubious also because these two quantities have
vastly different diffusivities, and therefore cannot be similarly distributed
in space. Magnetic filaments and sheets produced by the small-scale dynamo in
the diffuse warm gas can have scales as small as a few parsecs \citep{S07},
and the strength of this turbulent magnetic field can be about $5\mkG$. The
large-scale magnetic field varies over scales of order 1\,kpc, consistent with
the turbulent diffusivity of $10^{26}\cm^2\s^{-1}$ and time scale
$5\times10^8\yr$. The diffusive length scale of cosmic rays, based on
the diffusivity of $D\simeq10^{28}\cm^2\s^{-1}$ and the confinement time in
the disc, $\tau\simeq10^7\yr$, is about $(2D\tau)^{1/2}\simeq1\kpc$,
similar to { the length scale} of the large-scale magnetic field. On these grounds, it is not
impossible that the energy densities of cosmic rays and the
\textit{large-scale\/} magnetic field vary at similar scales, but this would
be very implausible for the total magnetic field. Then equipartition arguments
may be better applicable to observations of external galaxies, where the
linear resolution is not better than a few hundred parsecs, than to the case
of the Milky Way.

\section{Discussion}\label{Disc}
The general picture emerging from our results is that cosmic rays and
magnetic fields are slightly anticorrelated at the relatively small,
sub-kiloparsec scales explored here
($n\cra=\text{const}$ is also a viable possibility).
Such an anticorrelation can result from statistical
pressure equilibrium {(i.e. a statistically constant total pressure)} in the ISM, where cosmic rays and magnetic fields make
similar contributions to the total pressure. An additional effect leading to
an anticorrelation is the increase in the synchrotron losses of relativistic
electrons in stronger magnetic field.

{Strictly speaking, this conclusion applies to regions for which we have analysed the data: high Galactic latitudes in the
Milky Way and the outer parts of M33. However,
it is likely that this result reflects general features of cosmic ray propagation.}

{Local energy equipartition (or pressure equality)
between cosmic rays and magnetic field would produce stronger fluctuations
of synchrotron intensity than those observed.}
However, equipartition between cosmic
rays and magnetic field cannot be excluded at larger scales of order
1\,kpc and greater.
\citet{HBX98} indirectly make a similar conclusion concerning loss of equipartition at small
scales from their analysis of the radio--far-infrared correlation in M31.

Since magnetic fields and cosmic rays have vastly different diffusivities,
and therefore, must vary at very different scales, any strong correlation
between them can hardly be expected at scales smaller than 1\,kpc.
Correlated (or rather anticorrelated) fluctuations can, however, arise from
such secondary processes as the adjustment to pressure equilibrium, etc.

Our arguments and conclusions are based on observations and modelling of
synchrotron emission, a tracer of the electron component of cosmic rays. Thus,
our conclusions strictly apply only to the cosmic ray electrons.
However, the only significant difference
between the behaviour of electrons and protons in this context is that the
former experience higher energy losses due to synchrotron emission
and inverse Compton scattering off the relic microwave photons. The energy
loss time scale $\simeq4\times10^8\yr(E/1\GeV)^{-1}$ in a magnetic field of
$5\mkG$ in strength, for particles emitting at wavelengths larger than 1\,cm,
is much longer than the confinement time $10^7\yr$, so the energy losses are
negligible unless the local magnetic field is unusually strong. Therefore, we
extend our conclusions derived from analysis of synchrotron intensity fluctuations to
cosmic rays as a whole. Moreover, energy losses can only make the distribution
of the electrons more inhomogeneous than that of the protons, so that our
conclusions are robust with respect to this caveat.

{Our model, data and their analysis arguably match each other in the level of detail.
We do not include any latitudinal variation of the path length $L$
and the variation of the angular size of the turbulent cells, $l_0/L$ with Galactic
latitude  in the Milky Way. Instead, we restrict our analysis to the range
$|b|>30^\circ$ within which both $l_0/L$ and $L$ vary by a factor of two. The important
parameter, the square root of the number of turbulent cells along the path length,
$N^{1/2}\simeq(L/2l_0)^{1/2}$ then varies by a factor of about 1.5.
Since there are other parameters varying with galactic latitude (e.g., the
magnitudes of the random and regular magnetic fields, cosmic-ray intensity, etc.),
including the dependence of the path length and the correlation scale into the model would
make it significantly more complicated, if possible at all. Therefore, we prefer,
instead, to present our results in the form of plausible ranges that allow for the
numerous effects that remain beyond the framework of the model.}

{To simplify analytical calculations, we have adopted $s=-1$ for the synchrotron spectrum,
so that the synchrotron emissivity $\varepsilon$ is proportional to $B_\perp^{1-s}=B_\perp^2$.
We have verified that numerical results obtained with the more commonly used value $s=-0.7$,
where $\varepsilon\propto B_\perp^{1.7}$, differ insignificantly from those with  $s=-1$.}

{Our model includes magnetic field and cosmic ray distributions represented by a wide
range of scales, with the magnetic energy spectrum given by Eq.~\eqref{Mss}. However, the
spectral index of magnetic fluctuations only appears in the expressions
for the r.m.s.\ magnetic field fluctuations, Eq.~\eqref{sigmabs}, and the magnetic correlation
length, Eq.~\eqref{Bcorr}, through which it affects the number of correlation cells $N$.
Otherwise, the standard deviation of the synchrotron intensity is not sensitive to the
magnetic spectrum.}

{We have adopted $l_0=50$--$100\p$ for the correlation scale
of the random magnetic field. Estimates of the turbulent scales in the
magnetoionic medium of the Milky Way are numerous and divergent.}
\citet{RS77}
{discuss in detail techniques
for estimating turbulent scales from pulsar $\RM$ data and obtain
$l_0=100$--$150\p$ without any restrictive assumptions  regarding the
correlation between the fluctuations in magnetic field and thermals electrons.}
\citet{RK89}
{estimate the size
of a turbulent cells as $2l_0=55\p$ (using our notation) from the Faraday
rotation measures of pulsars. In fact, their result refers to the
size of the correlation cell of $\RM$ fluctuations and these authors
do not discuss how it is related to the correlation scale of magnetic
field; this relation depends on the degree of (anti) correlation between
the fluctuation in magnetic field and thermal electron density}
\citep{BSSW03}.
\citet{OS93}
{estimate the scale of magnetic field fluctuations from the
$\RM$ of close pairs of pulsars to obtain $2l_0=10$--$100\p$. Their model
includes fluctuations in thermal electron density but they are, presumably,
considered to be uncorrelated with magnetic field fluctuations; this assumption
can significantly affect the result.}
\citet{HBGM08}
{obtain the integral
(correlation) scale
from $\RM$ and depolarization of extragalactic radio sources;
their sample probes the inner Galaxy avoiding its central part.
These
authors obtain $l_0=1$--$5\p$ from the Faraday rotation measures and
$l_0=3.5$--$8.7\p$ from depolarization. The authors attribute the difference
from other estimates of the outer scale to a correspondingly smaller energy
input scale of interstellar turbulence of a few parsecs. Perhaps more plausibly,
the fluctuations in $\RM$, depolarization or any other parameter can
have a hierarchy of characteristic scales due, say, to interstellar shocks,
intermittent small-scale magnetic filaments, etc., and different methods can
be sensitive to only some of them.}
\citet{FBSBH11}
{deduce the correlation
scale of $\RM$ fluctuations from high-resolution observations of M51
by comparing the scatter in the values of $\RM$ observed under various
degrees of spatial smoothing and assuming that the standard deviation of $\RM$
scales as $l_0^{1/2}$ as predicted by theory} \citep[e.g., ][]{1998MNRAS.299..189S}.
{The resulting scale of $\RM$ fluctuations is $l_\mathrm{RM}=50\p$.} \citet{Houde:2013} {analysed the dispersion of synchrotron polarisation angles in high-resolution observations of M51 to estimate $l_0=98\pm5\p$ and $l_0=54\pm3\p$ parallel and perpendicular to the local mean-field direction respectively.}

{We discuss the relation between the correlation lengths of synchrotron intensity and
magnetic field in Section~\ref{scale}; this discussion and conclusions apply
to other observables as well. It is important that there is no
universal relation between the correlation lengths of, say, $B_\perp^2$ and $B$:
to establish such a relation, one has to know the auto-correlation functions of $B_\perp$
and $B_\parallel$. In addition, such observables as Faraday rotation measure,
total or polarized radio intensity
involve not only magnetic field but also number densities of thermal or relativistic
electrons. The cross-correlation between these variables and magnetic field are
also required to deduce the statistical properties of magnetic field.}
{ The comprehensive statistical analysis is recently suggested by \citet{Lazarian2012ApJ}. However the theoretical predictions discussed there is hardly possible to compare with available observational data. Only the simplest statistical characteristics give robust results.}

Our results can significantly change the interpretation of high-resolution
radio observations of the Milky Way and spiral galaxies. Present
interpretations, aimed at estimating the strength and geometry of interstellar
magnetic fields, rely heavily on the assumption of local equipartition between
cosmic rays and magnetic fields, at a scale corresponding to the resolution of
the observations. This assumption is acceptable if the resolution is not finer
than the diffusion length of the cosmic rays, about {say,} $1\kpc$. However, this
assumption is questionable when applied to observations at higher resolution.
We suggest a different procedure to interpret such observations. The
original total intensity radio maps should first be smoothed to the scale of the cosmic ray
distribution, 1\,kpc, where the equipartition assumption \textit{may\/} apply,
and the distribution of cosmic rays can be deduced from the smoothed data.
{(The smoothing length may depend on the local environment e.g. star formation rate, magnetic field etc. --- this requires further investigation using suitable cosmic ray propagation models.)}
After that, this distribution of cosmic rays can be used to deduce the magnetic
field distribution from the data at the original resolution. Since a larger
part of the synchrotron intensity fluctuations observed will be attributed to magnetic
fields, it is clear that this procedure will result in a more inhomogeneous
magnetic field than that arising from the assumption of local equipartition so
often used now.

\section*{Acknowledgments}
{We thank an anonymous referee for a careful and insightful report on the submitted version of the paper, which resulted in many improvements. We are also grateful to Wolfgang Reich for helpful comments.} Financial support from
the Royal Society and Newcastle University is gratefully acknowledged.
At various stages, this work was supported by the Leverhulme Trust
under Research Grant RPG-097, the Royal Society, and by the National Science Foundation under
Grant NSF PHY05-51164. RS acknowledges support from the grant YD-520.2013.2
of the Council of the President of the Russian Federation. Numerical simulations were partially performed on the supercomputer "URAN" of Institute of Mathematics and Mechanics
UrB RAS.

\appendix
\section{{The mean and standard deviation of the synchrotron intensity}}\label{app}

Here we derive analytical expressions for magnitude of relative synchrotron intensity fluctuations
$\delta_I=\sigma_I/I_0$, with $\sigma_I$ the standard deviation of the synchrotron
intensity $I$ and $I_0$, its mean value.
Relations between the statistical characteristics of synchrotron intensity and synchrotron emissivity $\varepsilon=n\cra
B_\perp^2$ are given by Eqs.~(\ref{I0e}) and (\ref{sigmaIe}).
Here we calculate $\aver{\varepsilon}$ and $\sigma_\varepsilon^2=\aver{\varepsilon^2}-\aver{\varepsilon}^2$ using
the cosmic ray model, partially correlated with magnetic fields, introduced in Section~\ref{Cray}.
Here overbar denotes ensemble averaging.
The calculations are quite straightforward although somewhat cumbersome.


We start with calculating $\aver{\varepsilon}$ using $n\cra$ from Eq.~\eqref{ncrafull}:
\begin{eqnarray*}
\aver{\varepsilon}&=&\aver{n\cra B_\perp^2}\\
	&=&n_0\aver{B_\perp^2}+\frac{\sigma_n C}{\sigma_{B^2}}(\aver{B^2 B_\perp^2} - \aver{B^2}\,\aver{B_\perp^2})\\
&&\mbox{}+\frac{\sigma_n \sqrt{1-C^2}}{\sigma_{F}}(\aver{F B_\perp^2} - \aver{\aver{F} B_\perp^2}).
\end{eqnarray*}
The last term vanishes since $F$ and $B_\perp^2$ are uncorrelated by construction, and hence
$\aver{F B_\perp^2}=\aver{F}\,\aver{B_\perp^2}$ and $\aver{\aver{F}\,B_\perp^2}=\aver{F} \aver{B_\perp^2}$.
For the same reason, all terms in $\aver{\varepsilon^2}$ that contain $F$ also vanish.

As each Cartesian vectorial component of the random magnetic field $\vec{b}$
is assumed to be a Gaussian random variable with zero mean value, $\aver{b}_i=0$,
we have, for the higher statistical moments,
\begin{equation}\label{shizo}
\aver{b_i^{2k}}=\frac{(2k)!}{2^k k!}\sigma_b^{2k},
\quad
\aver{b_i^{2k+1}}=0,
\quad
k=1,2,\ldots,
\end{equation}
where $i=x,y,z$. This allows us to calculate the higher-order moments
that contribute to $\aver{\varepsilon}$ and $\aver{\varepsilon^2}$ as follows:
\begin{eqnarray*}
\aver{B_\perp^4}&=&B_{0\perp}^4+\frac{8}{3} B_{0\perp}^2 \sigma_b ^2+\frac{8}{9} \sigma_b ^4,\\
\aver{B^4}			&=&B_{0\perp}^4+\frac{10}{3} B_{0\perp}^2 \sigma_b ^2+\frac{5}{3} \sigma_b ^4,\\
								&&\mbox{}+2 B_{0\perp}^2 B_{0\parallel}^2+\frac{10}{3}B_{0\parallel}^2 \sigma_b^2+B_{0\parallel}^4\\
\aver{B^2 B_{\perp}^4}&=&B_{0\perp}^4 B_{0\parallel}^2
        									+\frac{8}{3} B_{0\perp}^2 \sigma_b ^2 B_z^2
       										 +\frac{8}{9} \sigma_b ^4 B_{0\parallel}^2 \\
											&&\mbox{}+B_{0\perp}^6+\frac{19 }{3}B_{0\perp}^4 \sigma_b ^2
  												 +\frac{80}{9}B_{0\perp}^2 \sigma_b^4+\frac{56 }{27}\sigma_b ^6,\\
\aver{B^4B_{\perp}^4}&=&2 B_{0\perp}^6 B_{0\parallel}^2+14 B_{0\perp}^4 \sigma_b^2 B_{0\parallel}^2
													+B_{0\perp}^4 B_{0\parallel}^4\\
   									&&\mbox{}+ \frac{64}{3} B_{0\perp}^2 \sigma_b ^4 B_{0\parallel}^2
   												+\frac{8}{3} B_{0\perp}^2 \sigma_b^2 B_{0\parallel}^4\\
										&&\mbox{}+\frac{16}{3} \sigma_b^6 B_{0\parallel}^2+\frac{8}{9} \sigma_b ^4 B_{0\parallel}^4
   												+B_{0\perp}^8+\frac{34 }{3}B_{0\perp}^6 \sigma_b ^2 \\
										&&\mbox{}+\frac{109}{3} B_{0\perp}^4 \sigma_b^4 +\frac{104 }{3}B_{0\perp}^2\sigma_b^6
													+\frac{56}{9} \sigma_b^8.
\end{eqnarray*}
{The algebra involved in deriving these relations is rather daunting; we used symbolic algebra
software to derive these relations.}

We finally have from Eqs.~(\ref{I0e}), (\ref{sigmaIe}) and (\ref{ncrafull}):
\begin{equation}
I_0
=\frac{n_0 N\corr_\varepsilon}{9}
        \left[4 C \sigma^{-1}_{B^2}\delta_n\sigma_b^4+6\sigma_b^2
                +3\left(4 C \sigma^{-1}_{B^2}\delta_n\sigma_b^2+3\right) B_{0\perp}^2\right],
\label{I0}
\end{equation}
and
\begin{align}
\sigma_I^2&=\frac{n_0^2 N\corr_\varepsilon^2}{81}
        \Bigl\{224 C^2 \sigma^{-2}_{B^2} \delta_n^2 \sigma_b^8-81 \delta_n^2 \left(C^2-1\right)B_{0\perp}^4+2\sigma_b ^4
                                                        \nonumber\\
  &\mbox{}+18\sigma_b^4 \left[51 C^2 \sigma^{-2}_{B^2} \delta_n^2 B_{0\perp}^4+36 C \sigma^{-1}_{B^2} \delta_n
           B_{0\perp}^2 -4 \delta_n^2(C^2-1)\right]  \nonumber \\
  &\mbox{}+12 C^2 B_{0\parallel}^2\frac{\sigma_b^2}{\sigma^{2}_{B^2}} \delta_n^2 \left(9 B_{0\perp}^4
  				 +24 \sigma_b^2 B_{0\perp}^2 +8\sigma_b^4\right)\nonumber\\
&\mbox{}+108B_{0\perp}^2 \sigma_b ^2\left[\left(C \sigma^{-1}_{B^2}\delta_n B_{0\perp}^2+1\right)^2-2\delta_n^2(C^2-1)
                \right] \nonumber\\
   &\mbox{}+ 144 C \sigma^{-1}_{B^2} \delta_n \left(9C \sigma^{-1}_{B^2} \delta_n B_{0\perp}^2+1\right)\sigma_b^6\Bigr\},
                                                        \label{sigmaI}
\end{align}
where
\begin{equation}
\sigma_{B^2}^2=\aver{B^4}-\aver{B^2}^2
						  =B_0^4+\frac{10}{3}B_0^2\sigma_b^2+3\sigma_b^4.
\label{eq:App:sigB2}
\end{equation}

\bibliographystyle{mn2eC}
\bibliography{biblio}

\label{lastpage}
\end{document}